\documentclass[11pt]{article}
\usepackage[a4paper, total={6.5in, 9.5in}]{geometry}

\usepackage{authblk}
\usepackage{natbib}
\usepackage{graphicx}
\usepackage{xcolor}
\usepackage{amssymb}
\usepackage[utf8]{inputenc}
\usepackage[ngerman,english]{babel}
\usepackage[T1]{fontenc} 
\usepackage[super]{nth}
\usepackage{comment}
\usepackage{float}
\usepackage{hyperref}
\usepackage{amsmath}
\usepackage{threeparttable}
\usepackage{multirow}
\usepackage[font=footnotesize,labelfont=bf]{caption}
\usepackage[labelfont=bf]{subcaption}

\usepackage{verbatim}

\DeclareRobustCommand{\rchi}{{\mathpalette\irchi\relax}}
\newcommand{\irchi}[2]{\raisebox{\depth}{$#1\chi$}} % inner command, used by \rchi

\title{%The Politics of Language Choice: How War %influences active shifting of language use on Twitter 
The Politics of Language Choice: How the Russian-Ukrainian War Influences Ukrainians' Language Use on Twitter
}

\author[1]{Daniel Racek \thanks{Corresponding author: Daniel Racek, daniel.racek@stat.uni-muenchen.de}}
\affil[1]{Institute of Statistics, Ludwig-Maximilians-University Munich, Germany}
\author[2]{Brittany I. Davidson}
\affil[2]{School of Management, University of Bath, United Kingdom}
\author[3]{Paul W. Thurner}
\affil[3]{Institute of Political Science, Ludwig-Maximilians-University Munich, Germany}
\author[4]{Xiao Xiang Zhu}
\affil[4]{School of Engineering and Design, Technical University of Munich , Germany}
\author[1]{Göran Kauermann}

\begin{document}
\maketitle

\begin{abstract}
The use of language is innately political and often a vehicle of cultural identity as well as the basis for nation building. Here, we examine language choice and tweeting activity of Ukrainian citizens based on more than 4 million geo-tagged tweets from over 62,000 users before and during the Russian-Ukrainian War, from January 2020 to October 2022. Using statistical models, we disentangle sample effects, arising from the in- and outflux of users on Twitter, from behavioural effects, arising from behavioural changes of the users. We observe a steady shift from the Russian language towards the Ukrainian language already before the war, which drastically speeds up with its outbreak. We attribute these shifts in large part to users' behavioural changes. Notably, we find that more than half of the Russian-tweeting users shift towards Ukrainian as a result of the war.
\end{abstract}

\section{Introduction}
%social media's importance 
%news

Social media is critically important in today's society \citep{saroj2020use,dwivedi2021social,wong2021use}. In recent years, it has played a key role in a number of political shifts and crises \citep{makinen2008social,sadri2018crisis}. While social media has been found to amplify all manners of misinformation, propaganda, populism, and xenophobia \citep{morozov2012net,zhuravskaya2020political,flamino2023political}, it can also serve as a mechanism to call for aid and as a source for live updates of major events unfolding \citep{sacco2015using, rogstadius2013crisistracker, allcott2017social, kaufhold2020mitigating}.
\\

In this article, we analyse language use of Ukrainian citizens on social media
before and during the Russian invasion of Ukraine (subsequently referred to as war), where after years of tensions and open aggression between Russia and Ukraine \citep{marples2021war}, on 24th February 2022, Russian forces began to invade and occupy parts of Ukraine \citep{bigg_2022}. At the time of writing, it has been estimated that the war has led to over 23,000 civilian casualties \citep{ohchr_ukraine_2023} and hundreds of billions of dollars worth of damage \citep{lamb_rebuilding_2022,world_bank_2023}. This has caused worldwide unrest, alongside 8.2 million Ukrainian refugees recorded across Europe and 5 million registered for temporary protection \citep{refugeestats, ratten2022ukraine}. 
\\

%use this as people documenting war on social media --> people are using social media during wars (potentially in English)
The war in Ukraine is also taking place in the digital era, with social media coverage documenting the horrific events in up to real-time. This provides a unique digital trace of many first-hand accounts of the war, as citizens are communicating among each other and to the public. This is generally known as crisis informatics, whereby social media data are utilized before, during, or after emergency events
for use cases such as disaster monitoring, management, and prevention \citep{sacco2015using,reuter2018social,jurgens2018effect,kaufhold2020mitigating,dwarakanath2021automated}. Recent studies have demonstrated that tweets can capture events of political violence  \citep{dowd2020comparing} and can help in monitoring and understanding intra-country conflicts \citep{steinert2022state}. 
%which may be useful for monitoring and potentially predicting future political instabilities before they occur. This is increasingly known as crisis informatics,
%whereby utilizing social media for real-world uses, such as disaster monitoring, management, and responses \citep{kaufhold2020mitigating, sacco2015using}. For example, \cite{dowd2020comparing} demonstrated that tweets could capture events of political violence during the Kenyan elections in 2017, where they manually assessed the overlap between events in their twitter dataset and conflict events. Other work analyzed twitter images to examine violence and protest dynamics. The authors find that these images are a helpful mechanism to monitor and understand intra-country conflicts \citep{steinert2022state}. 
\\

%Any communication requires the use of a common language
In our work, the language of a tweet is of particular interest. Notably,
%\textcolor{red}{Hence, akin to the above mentioned studies, we make use of the real-time nature of Twitter content, and focus on the languages used by those located in Ukraine. This is because} 
the use of language is inherently political. Languages can be the cause of conflict \citep{laitin2000language} and they are often incorporated in cultural and ethnic identity definition and are the basis for nation building and political change \citep{smagulova2006kazakh,wright_2012}. After the dissolution of the USSR, most post-soviet countries implemented new language laws in order to assert their original native language and build a new nation \citep{smagulova2006kazakh,pavlenko2008multilingualism}. In Ukraine, after independence, many people were considering themselves Russians by nationality or Ukrainian with Russian as their main language. With the Law on Languages (1989) and a 10-year plan for a gradual transition back to Ukrainian (1991), the government aimed to reverse those effects, but was only moderately successful in achieving this goal, as census results show \citep{marshall2002post,stebelsky2009ethnic,kulyk2018shedding}. Only more recently, with the Euromaidan protests and the Russian military intervention in Crimea and the Donbas, surveys between 2012 and 2017 show a consistent and substantial shift away from Russian ethnic and linguistic identification towards Ukrainian %(or Ukrainian-Russian) 
practice \citep{kulyk2018shedding}. Respondents note an increasing engagement with the Ukrainian language and are more supportive of Ukraine as a direct result of the military intervention. 
\\

We investigate language choice and tweeting activity on Ukrainian social media from January 2020 to November 2022 using over 4 million geo-tagged tweets from more than 62,000 different users.
In doing this, we study how Ukrainian citizens (and non-citizens living there) respond to their country being aggressively attacked and invaded by its direct neighbour they share a long history and language with, and how the use of language evolved before and during this war. Our study allows us to follow the same set of users and observe their (change in) behaviour over both the short- and longer-term as the war breaks out and continues to unfold on an individual level. %here potentially something about field experiment?
Hence, we are able to comment on recent news articles outlining shifts in language use from Russian to Ukrainian as a direct result of the war \citep{Guardian.06.03.2023,Multilingual.2022}. Moreover, we are able to monitor long-term language trends even before the war without the necessity of relying on small-scale surveys nor the infrequent censuses, of which the last one was conducted in 2001.
\\

More specifically, we study overall trends in the number of tweets in the three main languages (Ukrainian, Russian, English) over time. Second, we investigate how these trends translate to users' individual tweeting activity and if changes result from the in- and outflux of users, common in online communities \citep{dabbish2012fresh,panek2018effects, ransbotham2011membership}, or if they result from users changing their behaviour over time \citep{davidson2019evolution, eichstaedt2020tracking, dzogang2016seasonal}. We quantify the magnitude of both effects respectively. Third, we study if changes in users' tweeting activity originate from shifts between languages and quantify the magnitude of these shifts. Fourth and finally, we take a closer look at those users that switch from predominately tweeting in Russian to predominately tweeting in Ukrainian with the outbreak of the war.

\section{Results}

\subsection{Data Collection, Cleaning \& Processing} \label{results_data}
%For this work, we use two key datasets. First, we have a dataset from Twitter, 
We collected tweets from 9th January 2020 to 12th October 2022 using the 1\% real-time stream of the Twitter Sample API \citep{pfeffer2022sample}. 
%We restricted data collection to tweets containing of any geo-information using the Filter API \citep{pfeffer2022sample}. We then subsequently filtered the dataset to tweets from Ukraine ("UA" country tag) only, to obtain our sample of Ukrainian tweets, as commonly done for such studies \citep{hu2020understanding}. Our sensitivity analysis indicates that through this two-stage filtering process, we were able to recover almost all Ukrainian tweets during this time period (see section \ref{sensitivity_analysis}).
During collection, we filtered the data such that we only gathered tweets containing geo-information using the Filter API. We then manually filtered the dataset to only retain tweets from Ukraine (denoted by the "UA" country tag), %This ensured our sample contained only Ukrainian tweets,
as common in the literature \citep{hu2020understanding}, and exclude any retweets.
Our subsequently conducted sensitivity analysis shows that through this two-stage filtering process, we were able to recover almost geo-tagged tweets from Ukraine during this time period (see section \ref{sensitivity_analysis}).
%We conducted a sensitivity analysis, which indicated that through this two-stage filtering process, we were able to recover almost all Ukrainian tweets during this time period (see section \ref{sensitivity_analysis}).
\\

We conducted an extensive spam filtering scheme, in which we 1) removed any duplicate tweets, 2) identified and removed potential spam bots by training a bot detection model following \cite{yang2020scalable}, 3) removed users with >100 tweets per day, 4) only kept tweets coming from official Twitter clients or Instagram, and 5) applied additional filtering rules specific to our dataset. This reduced our dataset from originally 4,453,341 tweets (62,712 users) down to 2,845,670 tweets (41,696 users). For an extensive description and rationale see section \ref{cleaning_and_preprocessing}.
\\

Unsurprisingly, social media is popular in Ukraine, particularly among the younger generation, with almost all citizens aged 18-39 in 2021 reporting that they use social media. For Twitter, user statistics are as follows: 18-29 (13\% usage), 30-39 (8\%), 40-49 (7\%), 50+ (1\%) \citep{statista_most_2022}. Hence, our subsequent findings are not necessarily applicable to the entire population. However, they still provide valuable insights into the language use of Ukrainians aged 18-49.

\begin{comment}
\subsubsection{Sentiment Classification}
We classified all remaining tweets according to their text sentiment into one out of three sentiment classes, \emph{negative}, \emph{neutral}, or \emph{positive}. For this, we employed the sentiment version of the large multilingual language model XLM-T \citep{barbieri2022xlm}. The model was built from a pre-trained XLM-R \citep{conneau2019unsupervised}, a multilingual version of RoBERTa \citep{liu2019roberta}, and then trained on 198M Tweets across many different languages (including Ukrainian and Russian). The sentiment version was then fine-tuned specifically for the sentiment classification task on 24k Tweets. As recommended, we preprocess the texts of tweets by replacing all user mentions by a user token and all URLs by a http token.
\end{comment}

\begin{comment}
\subsubsection{Topic modelling}
\textcolor{red}{potentially here if needed?}
\end{comment}

\subsection{Descriptive Findings}
%Our final cleaned Twitter sample consists of 2,845,670 tweets from 41,696 different users. The language distribution of these tweets is shown in \autoref{fig:lang_distr}. The language distribution of all collected tweets is shown in \autoref{fig:lang_distr}. We use the language field provided by the Twitter API, as generally employed in the literature \citep{mosleh2021cognitive,barbieri2022xlm}.
To determine the language of a tweet, in accordance with the literature \citep{mosleh2021cognitive,barbieri2022xlm}, we utilize the language field provided by the Twitter API.  Ukrainian (35.8\%) and Russian (35.4\%) tweets are most prevalent in our dataset, followed by English (11.5\%). A large proportion of tweets (11.1\%) is labeled as "undefined", which among others consists of tweets that are too short, contain only hashtags, or only have media links. All other languages have shares of 1.2\% or less. For the subsequent analysis we focus on tweets coming from the three main languages (English, Russian, Ukrainian) and discard all remaining tweets. A full breakdown of the language distribution is reported in section \ref{language_distribution}.
%Hence, we group together the latter with the undefined tweets and label them as "other" for the subsequent analysis. 
\\

%The sentiment in our sample is made up of 27\% negative tweets, 47.2\% neutral, and the remaining 25.8\% are classified as positive.
\begin{comment}
\begin{figure}[h!]
    \centering
    \includegraphics[scale=0.25]{Figures/language_distribution.png}
    \caption{Relative distribution of the top 10 languages across the entire sample.}
    \label{fig:lang_distr}
\end{figure}

\begin{figure}[!htb]
     \centering
     \begin{subfigure}[b]{0.48\textwidth}
         %\hspace*{-2cm}
         \centering
         \includegraphics[scale=0.25]{Figures/language_distribution.png}
         \caption{Relative distribution of the top 10 languages across the entire sample.}
        \label{fig:lang_distr}
         \vspace{0.35cm}
     \end{subfigure}
    \hfill
    \begin{subfigure}[b]{0.48\textwidth}
         %\hspace*{-2cm}
          \centering
        \includegraphics[scale=0.33]{Figures/lang_over_time_paper.png}
        \caption{Daily number of tweets in the three most common languages. First vertical line denotes the first mention of Russian troops mobilizing and the second line denotes the start of the war.}
        \label{fig:daily_tweets}
         %\vspace{0.35cm}
     \end{subfigure}
    \caption{Language distribution of tweets.}
     \label{fig:lang_tweets}
\end{figure}
\end{comment}

\begin{figure}[h!]
    \centering
    \hspace*{-0.5cm}
    \includegraphics[scale=0.68]
    {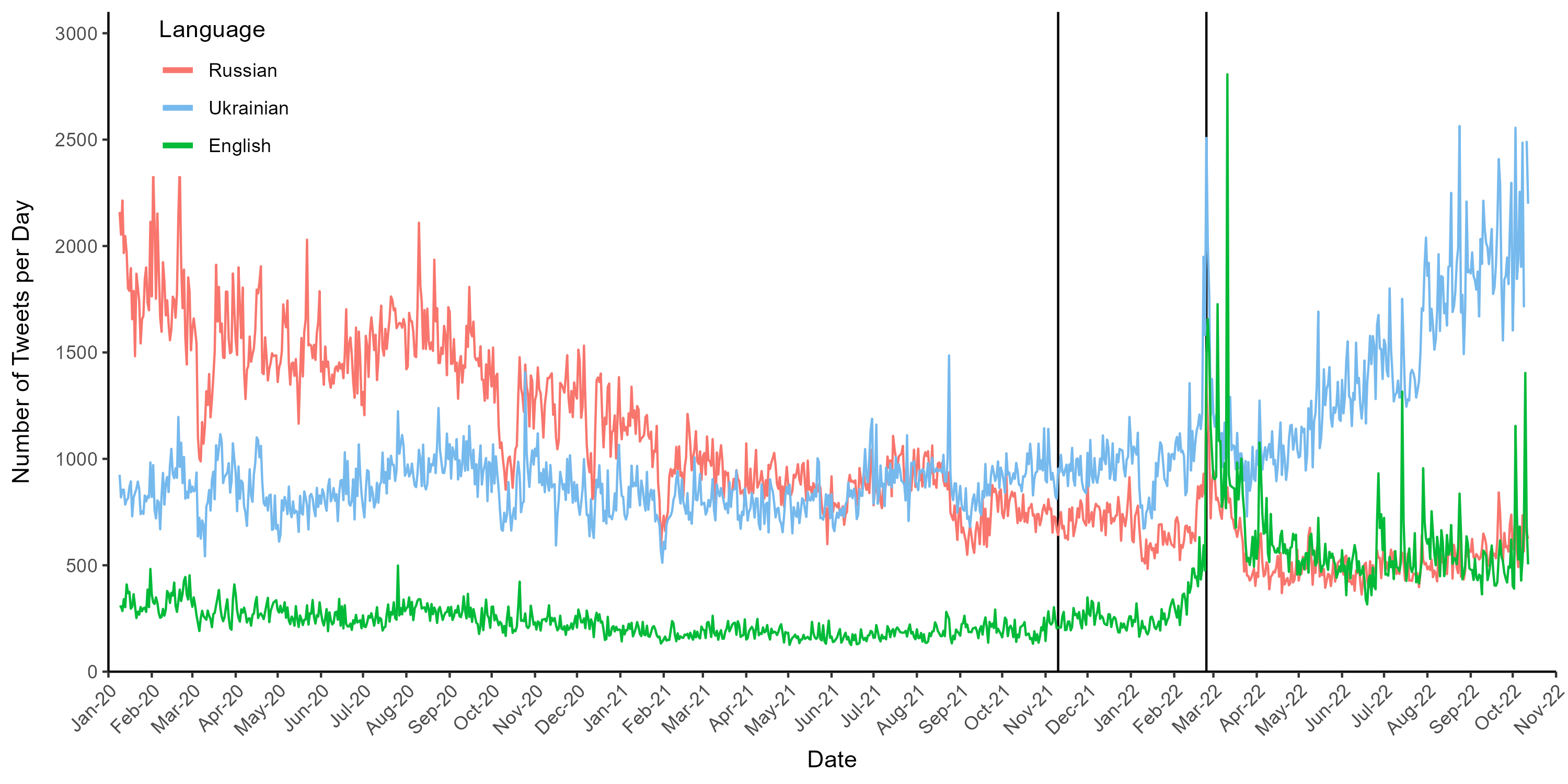}
    \caption{Daily number of tweets in the three most common languages (Russian, Ukrainian, English) from 9th January to 12th October (1,008 days). The first vertical line denotes the mobilization of the Russian troops along the Ukrainian border (11th November 2021). The second line denotes the outbreak of the war (24th February 2022).}
    \label{fig:daily_tweets}
\end{figure}

In our dataset, there are clear trends in the aggregate over time (\autoref{fig:daily_tweets}). 
In the beginning of 2020, we can see that Russian is the predominant language being used on Twitter in Ukraine, however, over time, this number gradually declines. The number of Ukrainian and English tweets on the other hand remains more or less constant over this initial time period. In the figure, we mark two key dates. On 11th November 2021, the United States officially report a mobilization of Russian troops along the Ukrainian border for the first time \citep{reuters_pentagon_2021,euronews_us_2021,ndtv_soldiers_2022}. We will subsequently call this the first signs of aggression. 24th February 2022 marks the begin of the Russian invasion of Ukraine (subsequently referred to as outbreak of the war). As we approach this outbreak, there is a clear spike in tweets across all three languages, with a larger spike in both English and Ukrainian. Afterwards, English and Russian remain mostly constant, although the former on a much higher level than before. For Ukrainian, there is a clear upward trend in the daily number of tweets after the outbreak of the war.
\\

Given these remarkable shifts in the number of tweets in the three considered languages, we want to investigate the underlying factors contributing to these changes. Note, that from the aggregate trends, we can not distinguish whether the observed patterns are due to large in- and outfluxes of users, which are common in online communities \citep{dabbish2012fresh,panek2018effects, ransbotham2011membership}, or whether the actively tweeting users change their behaviour over time \citep{davidson2019evolution, eichstaedt2020tracking, dzogang2016seasonal}. The disentanglement of this question is the aim of the rest of this article.

\subsection{User Activity} \label{user_activity}
In order to address this question, we restructure our dataset by aggregating the number of tweets made by each user in English (EN), Ukrainian (UA), and Russian (RU) in each week. (Note, that we employ the Ukrainian country code "UA" instead of the official Ukrainian language tag "UK" in order to avoid confusion.) This allows us to study users' individual behaviour over time. To obtain reliable results, we restrict the further analysis to users, who have tweeted in total at least ten times in any of the three languages. Furthermore, we choose weeks instead of days, as we are interested in general shifts and overall changes in behaviour over time, which are captured sufficiently well on a weekly basis. Through this weekly definition, we can dramatically reduce the size of our dataset, hence more complex modelling approaches become computationally feasible. We drop the first and last week in our dataset as these are incomplete (less than 7 days) and aggregate the remaining tweets on a weekly basis for each user and language. Finally within this, we are only considering weeks in which users are "active" (we define this as any week in which a user is tweeting at least once, as well as up to two weeks after), in order to account for the times in which users may be inactive for several weeks at a time or abandon their accounts. %(needed for disentanglement). 
Thus, our new sample ranges from 13th January 2020 to 10th October 2022 and consists of 143 analysis weeks, 13,643 users and 1,045,245 observations. 
\\

\begin{figure}[!htb]
    \centering
    \hspace*{-0.5cm}
    \includegraphics[scale=0.68]{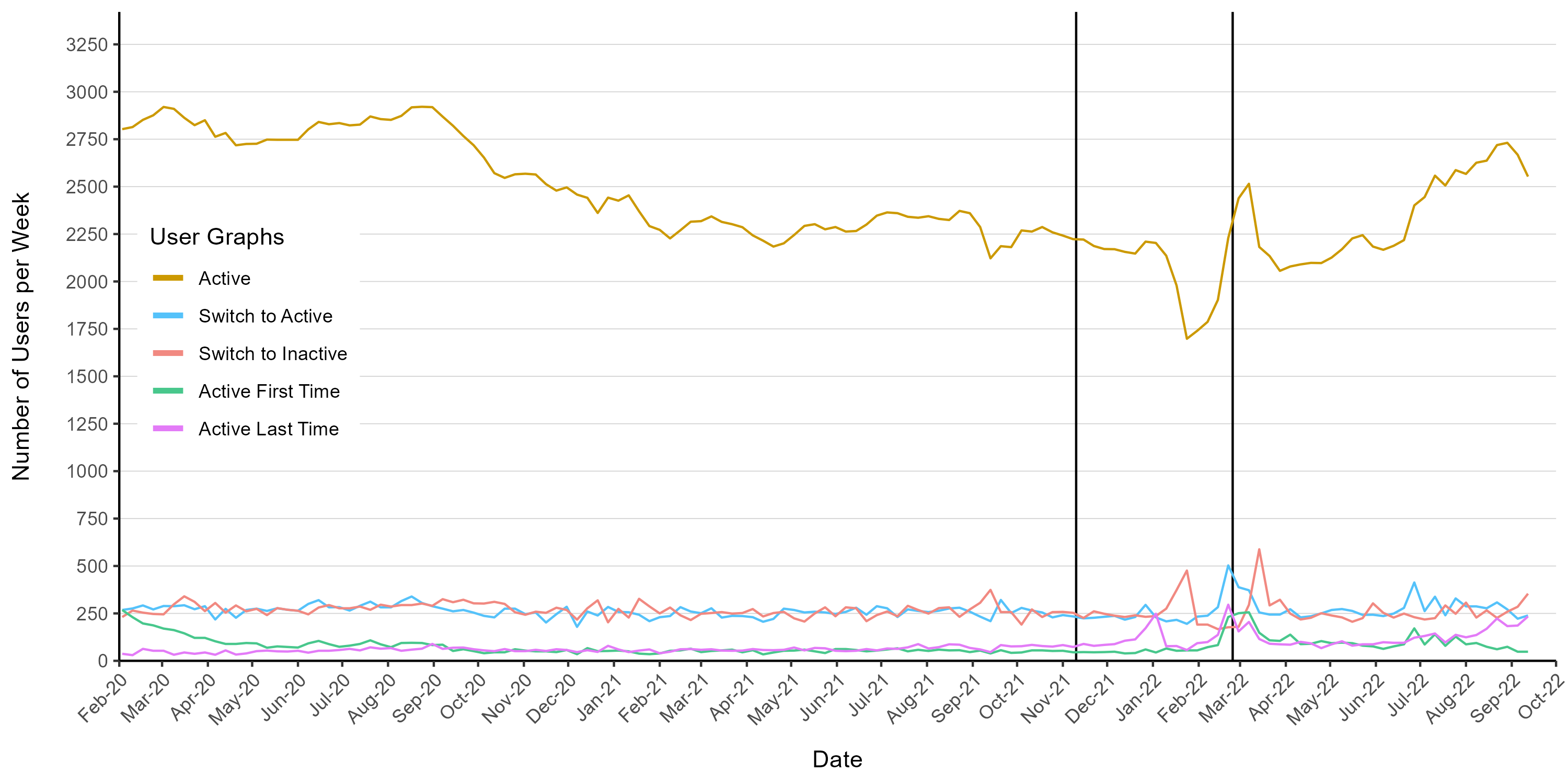}
    \caption{Weekly user activity graphs. The brown graph reports the number of active users in each week. The blue (red) graph reports the number of users who switch to active (inactive), the green the number of users who switch to active for the first time, the purple the number of users who were active for the last time, i.e. drop out of the sample altogether.
    All graphs, but particularly the latter two, are skewed upwards respectively downwards towards beginning and end of the study period due to the nature of how the dataset is constructed. Hence, we drop the first and last three weeks for visualization purposes (137 total weeks left). The full plot is available in supplementary material S.1. We also provide an additional version without the active user graph with a rescaling of the y-axis there. The first vertical line denotes the mobilization of the Russian troops along the Ukrainian border (11th November 2021). The second line denotes the outbreak of the war (24th February 2022).}
    \label{fig:user_activity}
\end{figure}

Using this definition of user activity, we can visualize the total amount of active users as well as turnover rates (switch from active to inactive and vice versa) over time (\autoref{fig:user_activity}). In the beginning of 2020, we have around 2,800 active users per week. This number gradually decreases to roughly 1,800 until we approach the outbreak of the war. Afterwards, the number of active users starts increasing again. Note the drop and subsequent spike in activity shortly before and with the outbreak of the war. Looking at the turnover rates, we find that there is a constant stream of $\sim$ 250 (potentially different) users per week that switch from active to inactive and vice versa. The aforementioned spikes are also evident in these turnover rates. Finally, we find that there are roughly 50 users per week that join our sample for the first time and about the same amount that leave it altogether. %\footnote{Note, that this only means these users start/stop tweeting with Ukrainian geo-information. Hence, they enter/drop out of our sample completely.} 
Both of these numbers almost double after the outbreak of the war. %, whereas the numbers of general activity switches stay mostly consistent before and after the war (with the exception that the number of active switches is slightly above the inactive switches).

\subsection{Tweeting Activity}
\begin{figure}[!htb]
     \centering
     \begin{subfigure}[b]{1\textwidth}
         %\hspace*{-2cm}
         \centering
         \includegraphics[scale=0.62]{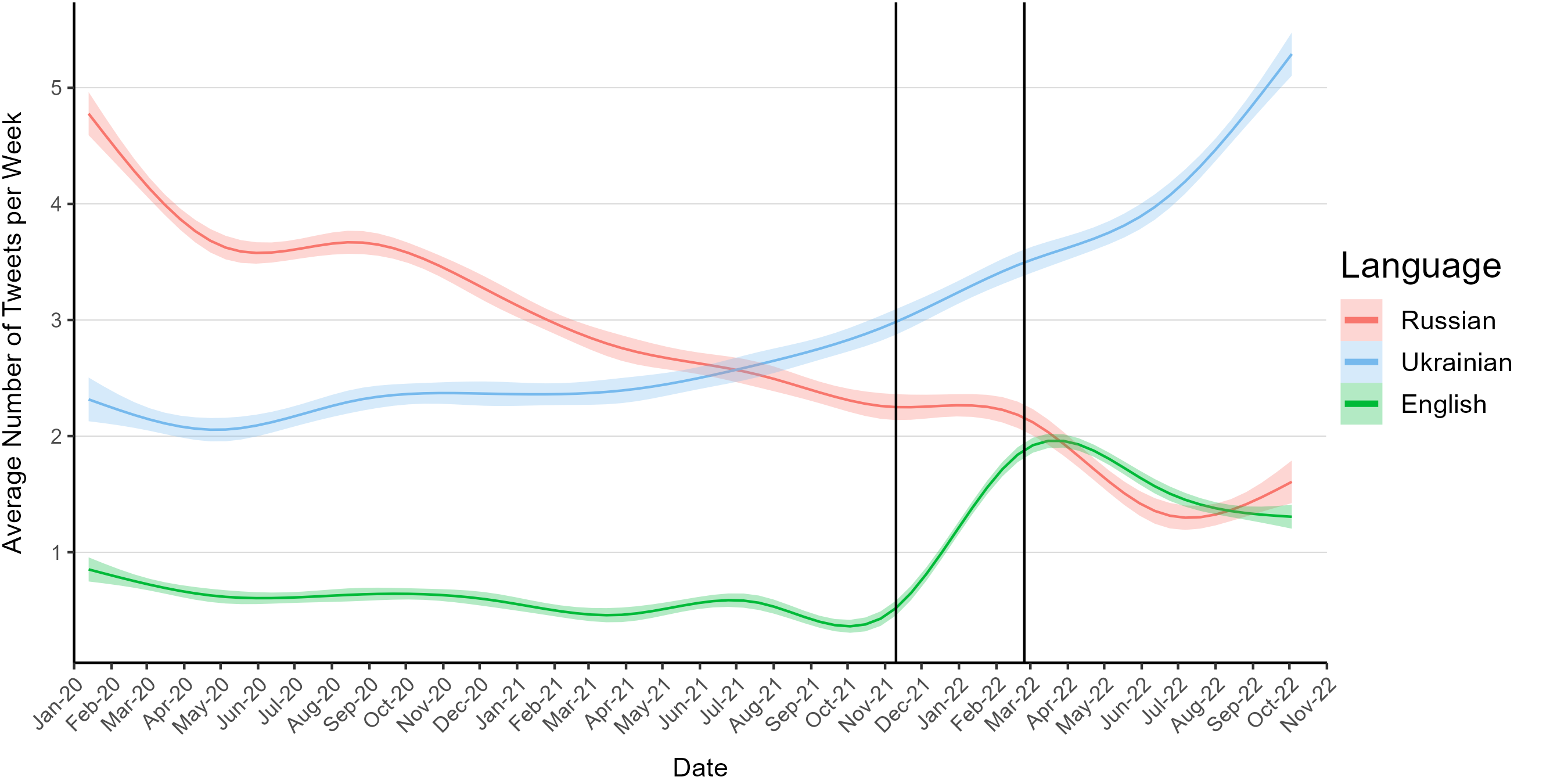}
          \caption{Average number of tweets. The graphs report a smoothed average of the published number of tweets per user in each week in each language. The shaded area depicts the 95\% confidence interval of the smooth fit. The non-smoothed version of the plot is available in supplementary material S.2.}
          \label{fig:avg_posts_per_user_actual_smoothed} 
         \vspace{0.35cm}
     \end{subfigure}
    %\hfill
     
     \begin{subfigure}[t]{0.48\textwidth}
     
         \hspace*{-1.5cm}
         \centering
         \includegraphics[scale=0.55]{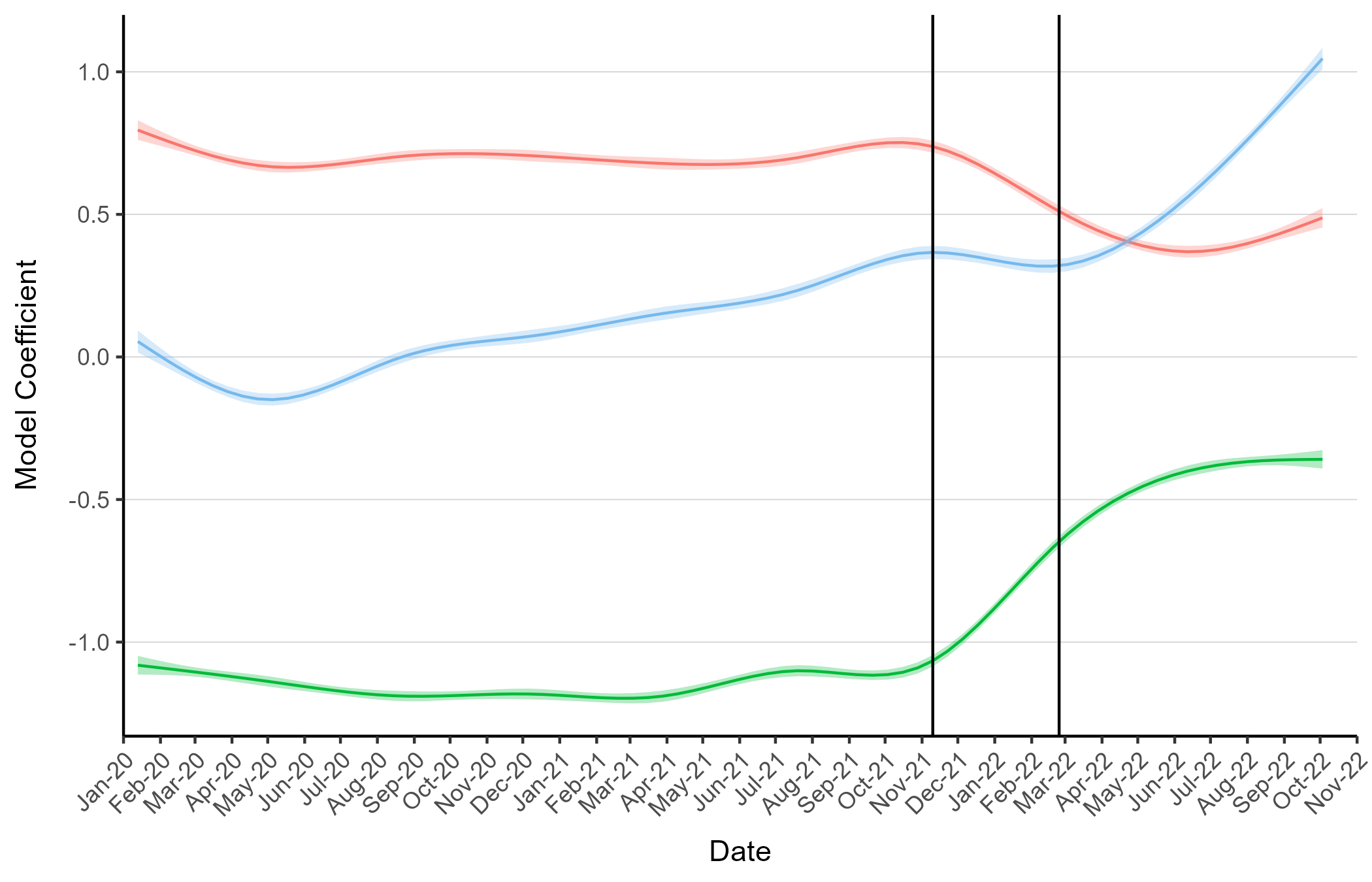}
         \caption{Sample effects. The graphs report a smoothed average of the random effects of the active users in each week in each language. The shaded area depicts the 95\% confidence interval of the smooth fit. The non-smoothed version of the plot is available in supplementary material S.3.}
         \label{fig:mean_res_tweet_model}
         %\vspace{0.35cm}
     \end{subfigure}
    \hfill
    \begin{subfigure}[t]{0.48\textwidth}
         %\hspace*{-2cm}
          \centering
        \includegraphics[scale=0.55]{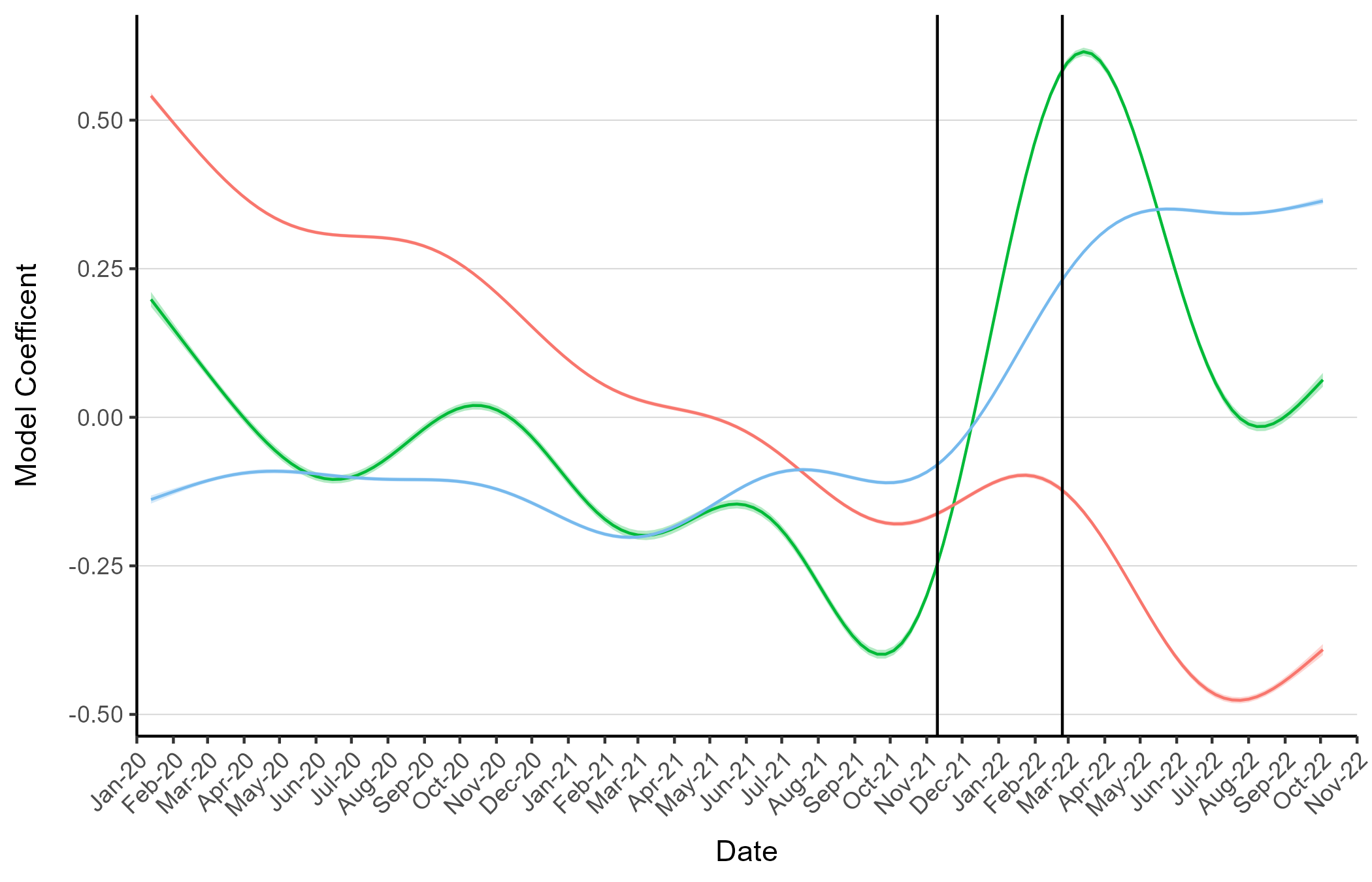}
        \caption{Behavioural effects. The graphs report the fitted global trend over all users in each week in each language. The shaded area depicts the 95\% confidence interval of the fitted effect.}
        \label{fig:smooth_tweet}
         %\vspace{0.35cm}
     \end{subfigure}
    \caption{Changes in the number of tweets per user. (a) visualizes the average number of tweets over time, (b) how sample changes affected the number of tweets, (c) how behavioural changes affected this number. The first vertical line denotes the mobilization of the Russian troops along the Ukrainian border (11th November 2021). The second line denotes the outbreak of the war (24th February 2022).}
     \label{fig:res_tweet_model}
\end{figure}

To obtain a better understanding on how the average active Ukrainian Twitter user changes over time, we visualize the average number of published tweets by a user in each language in \autoref{fig:avg_posts_per_user_actual_smoothed}. We smooth this average to highlight general trends. From the figure, we can clearly see that there are substantial shifts. Overall, the average number of RU tweets per user decreases constantly over time (from 4.8 to 2.2), the outbreak of the war being no exception. The average number of EN tweets decreases slightly until the war, where we notice a sudden uptick (from 0.5 to 1.9) followed by a steady decline. Meanwhile, the number of UA tweets slowly but steadily rises (from 2.4 to 3.0), with steeper increases after the first signs of aggression in November 2021 and no appearance of slowing down (5.3 at the end).
\\

By combining these findings with \autoref{fig:user_activity}, we can at least partially explain the aggregate trends evident in \autoref{fig:daily_tweets}. While the active user sample is shrinking over time, those users that stay (and join) the sample are tweeting more in UA. Hence, there is no decrease in the overall amount of UA tweets. We find the exact opposite for RU tweets. As the number of active users is declining, the users that stay active are tweeting less in RU, resulting in the visible decrease of aggregate RU tweets over time. Notably, so far, we do not know, if those changes in the average amount of tweets per user are simply driven by shifts in our active user sample (i.e., are those users that initially tweet a lot in RU leaving over time and this is why we see this decrease in the average?), or, if these changes are (at least partially) driven by behavioural changes in those users that remain active on Twitter (i.e., are the same users tweeting less in RU over time?). 
\\
\begin{comment}
\begin{figure}[!htb]
    \centering
    \includegraphics[scale=0.3]
    {Model_Figures/lang_shifts.png}
    \caption{Average number of tweets published by a user across languages over time. First vertical line denotes the first mention of Russian troops mobilizing and the second line denotes the start of the war.}
    \label{fig:lang_shift}
\end{figure}
\end{comment}

We address this through our tweet model described in section \ref{tweet_modelling}. We fit a generalized additive mixed model (GAMM) to predict the number of tweets made by each user in each language in each week, assuming a Poisson distribution. By incorporating both a smooth global time trend for each language, as well as user-specific random effects for each of the languages, we disentangle sample shifts (random effects) from behavioural changes (global trend). %Behavioural changes are captured by observing the same set of users over longer periods of time, while sample shifts can take place due to the in- and outflux of active users. 
Note, as on most other social media platforms, users have the option to create new accounts, which we cannot match to their prior ones. Hence, some of the behavioural effects might be underestimated and instead accounted for as sample effects.
\\

\autoref{fig:mean_res_tweet_model} visualizes the fitted average sample effects, i.e. the graphs depict how the average time-constant tweeting intensity in our active user sample changes over time due to user turnover. %This allows us to investigate possible shifts in the user sample over time. 
The figure shows, that the average RU tweeting intensity is mostly constant over time until November 2021, where aggression starts. From that point onward, in the span of only a few months, we see a decline of 22\% in RU tweets from November 2021 to October 2022 (end of study period), solely attributed to changes in the user sample during that period. For EN, we find somewhat of an opposite effect. Similarly, there are only minor fluctuations until November 2021. But afterwards, there is a sharp increase of 104\%. Taking a look at UA, we find a long-term increase of about 37\% before the aggression starts. This increase comes to a hold shortly before the war, and significantly speeds up in the weeks after (+97\%). All (relative) effect sizes calculated between the most relevant dates in our analysis period (start of study period, first signs of aggression, outbreak of war, end of study period) are reported in \autoref{tab:full_results_tweet_model}.
%First, let us take a look at our random effects (\autoref{fig:res_tweet_model}), i.e. investigate possible shifts in our user sample over time. \autoref{fig:mean_res_tweet_model} visualizes the average (smoothed) mean of our random effects (REs) over time. Hence, the figure describes how the average time-constant posting intensity of users changes over time due to changes in the sample. 
%From the figure we can see that the average Russian RE is mostly constant over time until the first signs of aggression. Then, in the span of only a few months we see a decline of 23\% in RU tweets from week 93 to week 144, solely attributed to changes in the user sample during that period. We find somewhat of an opposite effect for the English tweets. Here, again, there is almost no change in the average initially. Once the aggression starts, there is a sharp increase in EN tweets of 112\% that is exclusively attributed to the user turnover in this period. Finally, taking a look at UA tweets, we can find there is a long-term increase of about 34\% even before the war from week 2 to week 93. This increase comes to a hold shortly before the war, and significantly speeds up in the weeks after (+107\%). All (relative) effect sizes calculated between the most relevant weeks of our study period (start of sample, first signs of aggression, outbreak of war, end of sample) are reported in \autoref{tab:results_re_tweet_model}.
\\

\begin{table}[!htbp] 
\centering
\begin{threeparttable}
\caption{Tweet Activity Effect Sizes between Key Dates}
\fontsize{9.5}{11.5}\selectfont
\begin{tabular}{c c c c c} 
 \hline
\multirow{2}{*}{Language}  & \multicolumn{4}{c}{Sample Effects}   \\ 
\cline{2-5}
 &  Start - Aggression  & Aggression  - War   & War - End Study & Aggression - End Study\\
 \hline
English & +1.36\% & +51.45\% & +34.82\% &  +104.19\%\\ 

Ukrainian & +36.54\% & -4.99\% & +107.33\% &  +96.97\%\\ 

Russian & -5.42\% & -19.92\% & -2.92\% &  -22.26\%\\ 
\hline
& \multicolumn{4}{c}{Behavioural Effects}  \\
\hline
English & -36.75\% & +130.11\% & -39.98\% &  +38.09\%\\ 

Ukrainian & +5.71\% & +35.72\% & +15.184\% &  +56.32\%\\ 

Russian & -50.58\% & +4.68\% & -23.86\% &  -20.30\%\\ 
\hline
\end{tabular}
\caption*{Notes: Effect sizes for both sample and behavioural changes extracted from the tweet model described in section \ref{tweet_modelling} between key dates. All effect sizes are relative increases in the number of tweets between the two respective dates. Start: start of the study period---13th January 2020. Aggression: first official US report of a mobilization of the Russian troops along the Ukrainian border---11th November 2021. War: outbreak of the war---24th February 2022. End Study: end of the study period---10th October 2022.}
\label{tab:full_results_tweet_model}
\end{threeparttable}
\end{table}

Next, we will investigate behavioural changes using \autoref{fig:smooth_tweet}. The graphs depict how the tweeting behaviour of the active users changes throughout the study period, when controlling for the user turnover (sample effects). Starting with RU, we notice that users are tweeting less and less over time. From January 2020 to November 2021, users tweet 51\% less in RU due to behavioural changes. Subsequently, we see a small rise with the outbreak of the war (+5\%), followed up by an even steeper decline (-24\%). In contrast, UA is reasonably consistent in its use up until the start of aggression. From there, we observe a surge (+36\%) until the outbreak of the war, followed by a gentler increase (+15\%) after. Finally, looking more closely at EN tweeting behaviour, we can observe a general downward trend (-37\%) until November 2021. Once the aggression starts, there is a huge spike (+130\%), as users are tweeting a lot more in EN. After the outbreak of the war, this somewhat reverses (-40\%), however, without dropping back down to pre-aggression levels. A full breakdown of all changes is reported in \autoref{tab:full_results_tweet_model}. 
%Next, we will investigate behavioural changes (within user changes) using the smooth global time trends visualized in \autoref{fig:smooth_tweet}. Starting with Russian, we notice that users are tweeting less and less over time. From the beginning of the dataset until the first aggression (week 2 to 93), users tweet 51.33\% less in RU due to active behavioural changes. Subsequently, we see a small rise with the outbreak of the war (+6.29\%), followed by an even steeper decline (-23.86\%). In contrast, Ukrainian is reasonably consistent in its use up until the start of aggression, where there is a stark increase of 39.19\% until the outbreak of war, followed by a gentler increase (+15.18\%) after. Finally, looking more closely at English tweets, we see a general downward trend in tweeting behaviour (-43.91\%). Once the aggression starts, we can observe a huge spike with people tweeting a lot more in EN (+159.50\%). After the outbreak, this somewhat reverses, but without dropping back down to pre-aggression levels. A full breakdown of all changes is reported in \autoref{tab:results_smooth_tweet_model}. 
\\

\begin{comment}
\begin{table}[!htbp] 
\centering
\begin{threeparttable}
\caption{Average Behaviour Change Effect Sizes}
\fontsize{9.5}{11.5}\selectfont
\begin{tabular}{c c c c c} 
 \hline
\multirow{2}{*}{Language}  & \multicolumn{4}{c}{Change in Week Interval}   \\ 
\cline{2-5}
 &  Start - Aggression  & Aggression  - War   & War - End & Aggression - End   \\
 \hline
English & -36.75\% & +130.11\% & -39.98\% &  +38.09\%\\ 

Ukrainian & +5.71\% & +35.72\% & +15.184\% &  +56.32\%\\ 

Russian & -50.58\% & +4.68\% & -23.86\% &  -20.30\%\\ 
\hline
\end{tabular}
\caption*{}
\label{tab:results_smooth_tweet_model}
\end{threeparttable}
\end{table}
\end{comment}

Overall, we can conclude that there are only minor sample shifts pre-dating aggression that affected tweeting activity, but major shifts thereafter. In terms of behaviour, we can already see steady changes early on, which significantly intensify with the war. However, as of yet, we cannot exactly pinpoint where those changes come from. Are users that already tweet in UA simply tweeting more with the outbreak of the war, or is it possible that users are actively switching the language they are tweeting in?

\subsection{Choice of Language}

%A nuance we haven't investigated so far, is where some of those changes come from. 
%Are users simply tweeting more in their main language, or are they actively switching their language(s)? We can analyse this more closely with the language models proposed in section \ref{language_modelling}. In a similar fashion, to how we have analyzed the number of tweets made by users over time, we can now also analyze users' probability to tweet in a specific language (over another) over time. Are users simply tweeting more in their main language, or are they actively switching their language(s)?

\begin{figure}[!htb]
     \centering
     \begin{subfigure}[b]{1\textwidth}
         %\hspace*{-2cm}
         \centering
        \includegraphics[scale=0.62]{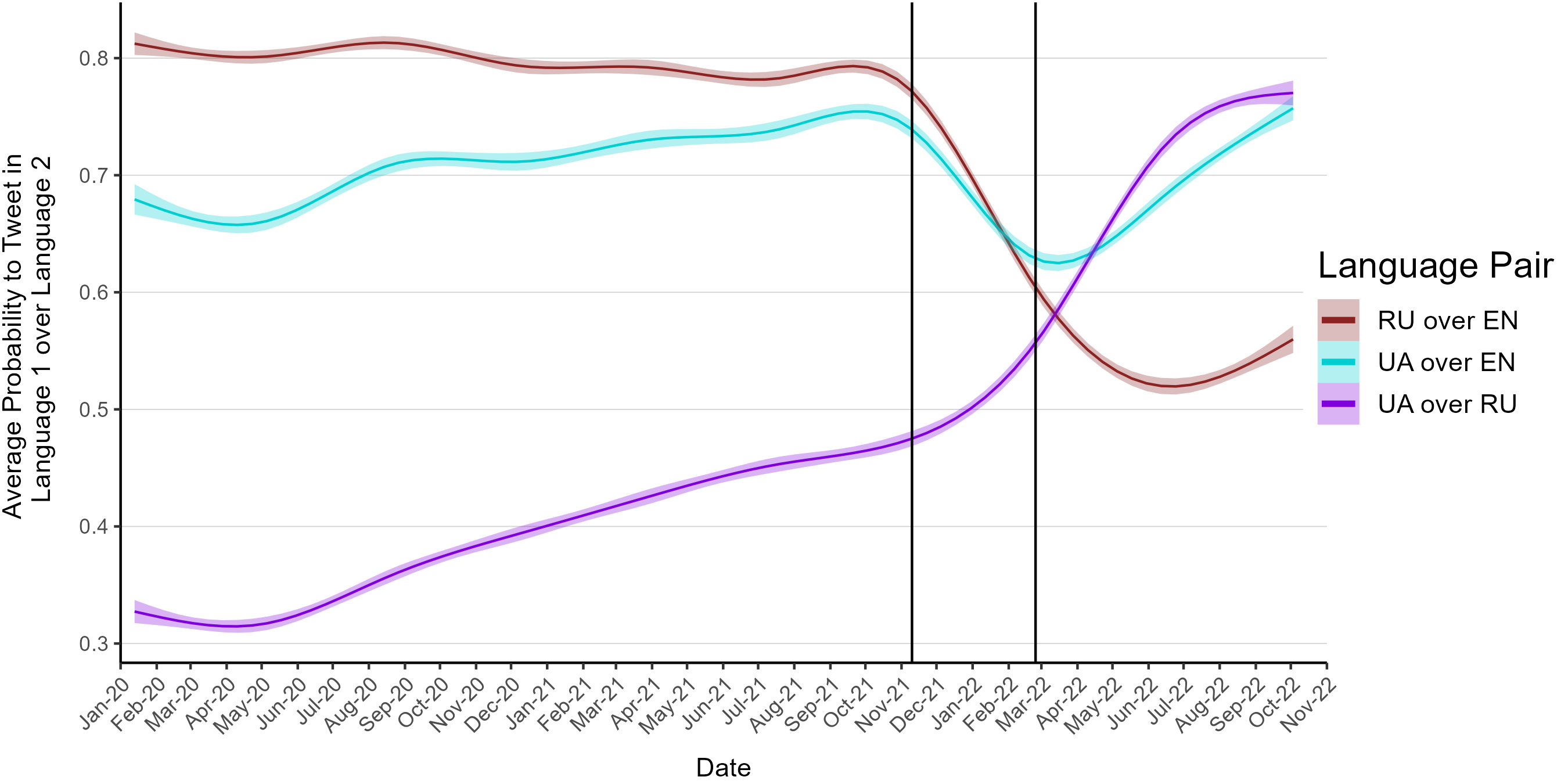}
        \caption{Average language probability. The graphs report a smoothed average of the probability to tweet in language one over language two per user in each week for the tree language pairs. The shaded area depicts the 95\% confidence interval of the smooth fit. The non-smoothed version of the plot is available in supplementary material S.2.}
    \label{fig:joint_actual_probabilites_smoothed}
     \vspace{0.35cm}
     \end{subfigure}
    %\hfill
    \begin{subfigure}[t]{0.48\textwidth}
         \hspace*{-1.5cm}
          \centering  
         \includegraphics[scale=0.55]{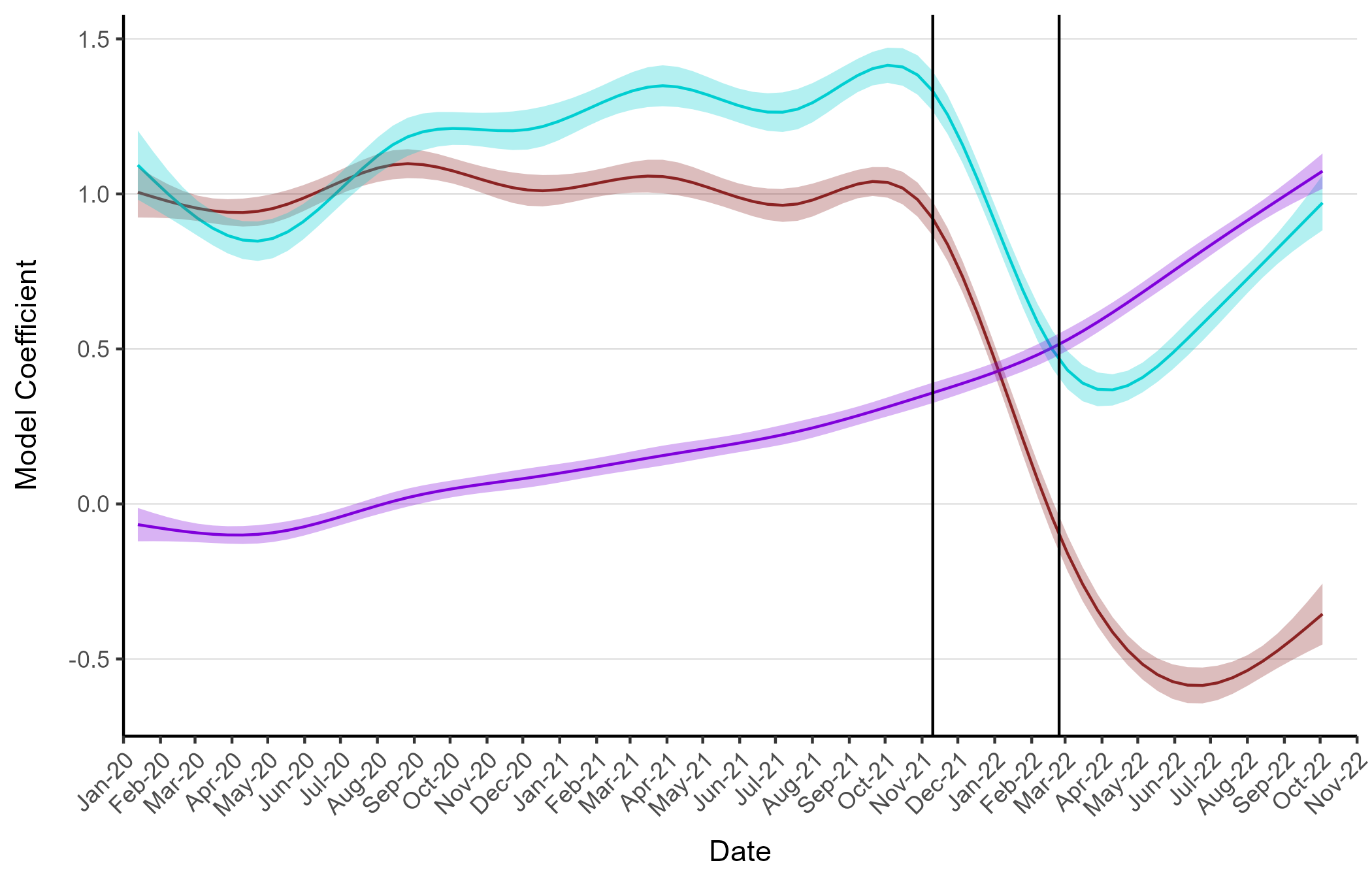}
         \caption{Sample effects. The graphs report a smoothed average of the random effects of the active users in each week for all three language pairs (hence for all three language GAMMs). The shaded area depicts the 95\% confidence interval of the smooth fit. The non-smoothed version of the plot is available in supplementary material S.3. }
         \label{fig:mean_res_language_model}
          %\vspace{-0.35cm}
     \end{subfigure}
    \hfill
    \begin{subfigure}[t]{0.48\textwidth}
     \centering
    \includegraphics[scale=0.55]{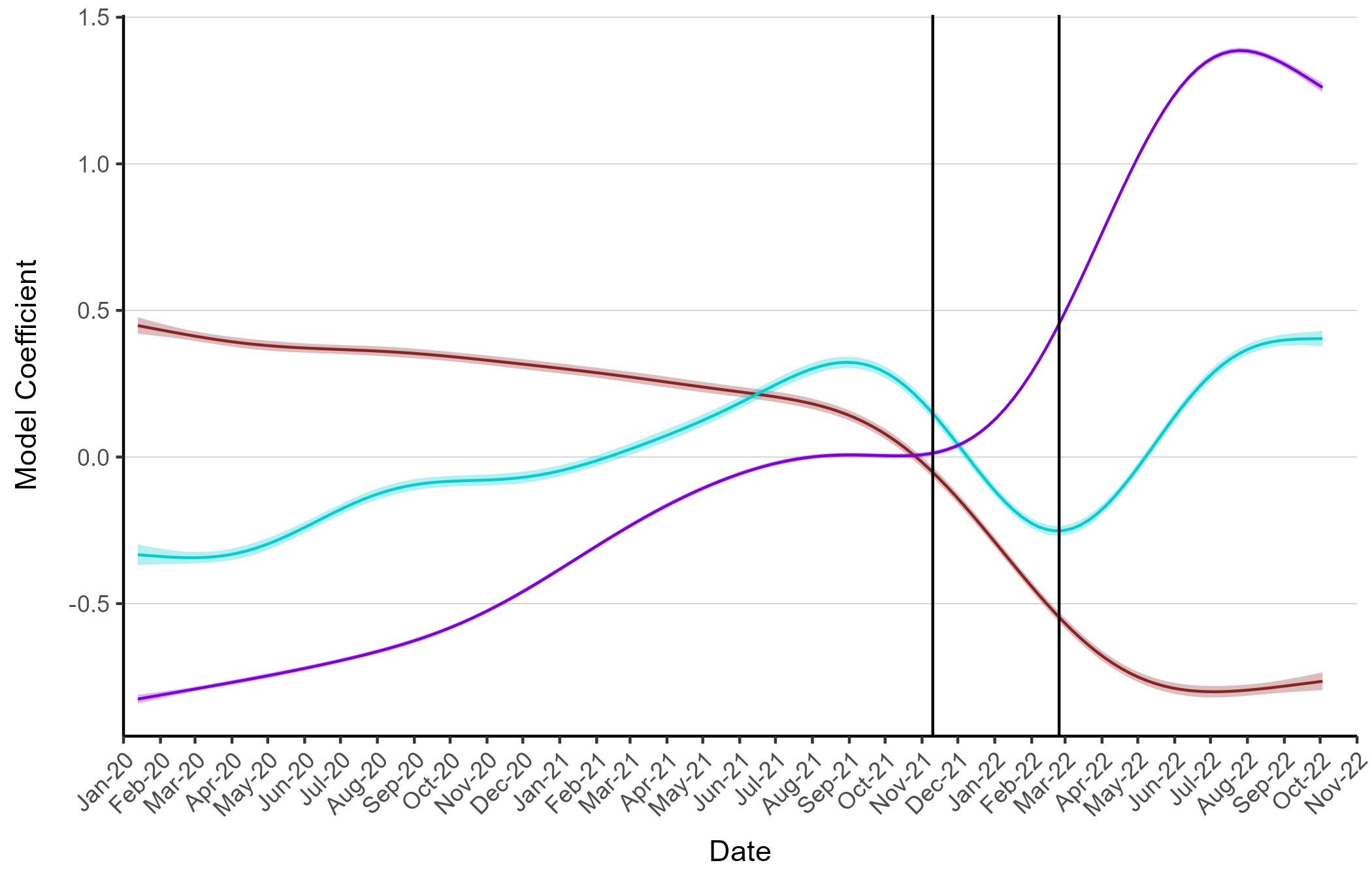}
    \caption{Behavioural effects. The graphs report the fitted global trend over all users in each week for all three language pairs (hence for all three language GAMMs). The shaded area depicts the 95\% confidence interval of the fitted effect.}
    \label{fig:smooth_lang}
     \vspace{0.3cm}
    \end{subfigure}
    \caption{Changes in the choice of language per user. (a) visualizes the average probability to tweet in one language over another, (b) how sample changes affected the probability, (c) how behavioural changes affected the probability. The first vertical line denotes the mobilization of the Russian troops along the Ukrainian border (11th November 2021). The second line denotes the outbreak of the war (24th February 2022).}
     \label{fig:res_lang_model}
\end{figure}

We analyze the choice of language more closely in the following. 
As we are interested in shifts between the individual languages, we look at the pairwise probability to tweet in one language over another over time. Hence, the probability reports how likely it is that a user tweets in language one (e.g. UA) over language two (e.g. EN).
With three languages, this pairwise evaluation gives us a total of three different language pairs (UA over RU, UA over EN, RU over EN), where the order in which we specify each pair is irrelevant. \autoref{fig:joint_actual_probabilites_smoothed} visualizes how these pairwise probabilities evolved for an average user over time.
%Hence, we can analyze how the average user changes his language preferences throughout the study period.
For RU over EN  the probability is mostly constant (80\% to tweet in RU) until aggression starts, from where it continuously drops down to 55\%. For UA over EN we see small increases over time (68\% to 74\%). With the mobilization of the Russian troops, we see a drop (63\%), followed by a rise back to pre-aggression levels months into the war. Finally, for UA over RU we see a completely different pattern. Initially, the probability to tweet in UA is low (33\%), from where it continues to rise consistently. In the weeks leading up to the war, there is a significant speed up in this shift, resulting in a probability of 77\% to tweet in UA over RU towards the end of the analysis period in October 2022.
\\

%\begin{figure}[!htb]
 %           \centering
 %       \includegraphics[scale=0.3]{Model_Figures/joint_actual_probabilities.png}
 %       \caption{Average probability of a user to tweet in one language (over another) over time. First vertical line denotes the first mention of Russian troops mobilizing and the second line denotes the start of the war.}
%\end{figure}

Similarly to before, we can disentangle sample shifts from behavioural changes through statistical modelling. In summary, we fit a GAMM to model users' pairwise language probability to tweet over time, assuming a binomial distribution. As before, we include a smooth global time trend and user-specific random effects into the model. We fit such a model, for all three aforementioned language-pairs. A full description is provided in section \ref{language_modelling}.
%In a similar fashion to before, we can disentangle sample and behaviour effects using the language models proposed in section \ref{language_modelling}. Here, we model users' probability to tweet in a specific language (over another) over time. In summary, we fit a GAMM to predict the probability a user tweets in language one (e.g. UA) over language two (e.g. RU), assuming a binomial distribution. We similarly incorporate a smooth global time trend and user-specific random effects into the model. We repeat this for each of our three language-pairs (UA vs. RU, UA vs. EN, RU vs. EN).
\\

\autoref{fig:mean_res_language_model} visualizes the fitted average sample effects across all three models, i.e. the graphs depict how the average time-constant tweeting probabilities in the active user sample change over time. As we are working with coefficients of a logistic regression, changes must be interpreted with respect to changes in the odds.
The figure shows that for RU over EN, initially, there are no relevant sample shifts (on average). However, as we approach the outbreak of the war, we can report a large drop in the odds, as users are 64\% less likely to tweet in RU over EN than before, with further decreases thereafter (-24\%). For UA over EN, we find a small to moderate increase until aggression (+29\%) due to sample shifts, followed by a large drop until war outbreak (-58\%),  which is recovered in the months after (+64\%). Finally, for UA over RU, there is a constant increase in the odds over time (+50\%), which significantly speeds up once aggression starts (+101\% until October 2022). \autoref{tab:full_results_lang_model} details all changes. 
%We will start with the sample changes, visualized by the average REs of each model over time in \autoref{fig:mean_res_language_model}. As we are working with coefficients of a logistic regression, changes must be interpreted with respect to changes in the odds (also known as odds ratio).\footnote{The odds ratio is defined as $odds = p/(1-p)$. Hence, it describes how likely an event is going to happen compared to not happen. In this setting, it describes how likely it is to tweet in language 1 over language 2.} From the figure, we can see that for RU over EN there were initially no relevant sample changes (on average). However, once the aggression starts, we see a large drop in the odds, and users are 74\% more likely to tweet in EN over RU than before. We find that for UA over EN there is a small to moderate increase until aggression (37\%) caused by sample changes, followed by a large drop until war outbreak (-58\%), which is quickly recovered after (+59\%). Finally, for UA over RU we see a constant increase in the odds over time (+44\%), which speeds up significantly once aggression starts (+107\%). \autoref{tab:results_re_lang_model} details all these changes. 
\\

Combining this with the results from the previous section, we can conclude that the user turnover in the first 1.5 years shifts the sample such that users are more likely to tweet in UA (than RU or EN), but not at the expense of either of the two other languages, as tweet levels are (mostly) steady for both. %This means, that users that are joining are tweeting more in UA, but at the same time are still continuing to tweet in both EN and RU. %\footnote{Note, that this does not mean each user is tweeting in all three languages.} 
As we approach the outbreak of the war, this drastically changes. Then, the user sample clearly shifts away from RU, as users are instead tweeting more in EN (initially) and UA (long-term). Upon further investigation (supplementary material S.4 and S.5), we find that users tweeting in RU start leaving around November 2021 (start of aggression), with EN users joining. The former continue to leave as the war unfolds, with some of the latter also slowly leaving the sample again over time. This is also reflected in the increase of the UA odds over time (UA over RU consistently, UA over EN as war continues).
%This changes with the first signs of aggression and the outbreak of the war. Then, we see a clear shift away from RU in the sample, as users are instead tweeting in EN (initially) and UA (long-term). Upon further investigation, we find that users tweeting in RU start leaving around week 93, with EN users joining. The former continue to leave as the war breaks out and continues, whereas some of the latter seem to slowly leave the sample again. This is also reflected in the increase of the UA odds over time (in UA over RU consistently, UA over EN as war continues to unfold). %\textcolor{red}{potentially add how we inferred this}
\\

%\begin{figure}[!htb]
%         \centering  
%         \includegraphics[scale=0.3]{Model_Figures/joint_res.png}
%         \caption{REs language model}
%         \label{fig:mean_res_language_model}
%\end{figure}
\begin{comment}
\begin{table}[!htbp] 
\centering
\begin{threeparttable}
\caption{Average Sample Change Effect Sizes - Language Models}
\fontsize{9.5}{11.5}\selectfont
\begin{tabular}{c c c c c} 
 \hline
\multirow{2}{*}{Language}  & \multicolumn{4}{c}{Change of Odds in Week Interval}   \\ 
\cline{2-5}
 &  Start - Aggression  & Aggression  - War   & War - End & Aggression - End   \\
 \hline
UA vs. RU & +49.58\% & +13.37\% & +77.01\% &  +100.68\%\\ 

UA vs. EN & +28.66\% & -58.47\% & +63.79\% &  -31.98\%\\ 

RU vs. EN & -6.66\% & -63.71\% & -24.36\% &  -72.55\%\\ 
\hline
\end{tabular}
\caption*{}
\label{tab:results_re_lang_model}
\end{threeparttable}
\end{table}
\end{comment}

\begin{table}[!htbp] 
\centering
\begin{threeparttable}
\caption{Language Choice Effect Sizes between Key Dates}
\fontsize{9.5}{11.5}\selectfont
\begin{tabular}{c c c c c} 
 \hline
\multirow{2}{*}{Language}  & \multicolumn{4}{c}{Sample Effects}   \\ 
\cline{2-5}
 &  Start - Aggression  & Aggression  - War   & War - End Study & Aggression - End Study\\
 \hline
UA over RU & +49.58\% & +13.37\% & +77.01\% &  +100.68\%\\ 

UA over EN & +28.66\% & -58.47\% & +63.79\% &  -31.98\%\\ 

RU over EN & -6.66\% & -63.71\% & -24.36\% &  -72.55\%\\ 
\hline
& \multicolumn{4}{c}{Behavioural Effects}  \\
\hline
UA over RU & +130.99\% & +52.08\% & +129.24\% &  +248.63\%\\ 

UA over EN & +63.41\% & -33.61\% & +92.663\% &  +27.90\%\\ 

RU over EN & -38.89\% & -38.69\% & -20.659\% &  -51.36\%\\ 
\hline
\end{tabular}
\caption*{Notes: Effect sizes for both sample and behavioural changes extracted from the language model described in section \ref{language_modelling} between key dates. All effect sizes are relative increases in the odds between the two respective dates. Start: start of the study period---13th January 2020. Aggression: first official US report of a mobilization of the Russian troops along the Ukrainian border---11th November 2021. War: outbreak of the war---24th February 2022. End Study: end of the study period---10th October 2022. }
\label{tab:full_results_lang_model}
\end{threeparttable}
\end{table}

\autoref{fig:smooth_lang} reports behavioural language changes across all three language pairs, when controlling for the user turnover. For RU over EN we see a constant decline in the odds over time (-33\% to tweet in RU), which further speeds up once aggression starts (-55\%). For UA over EN we see the exact opposite, as over time users are more likely to tweet in UA (+81\% in odds). This change reverses with the start of aggression and the outbreak of the war (-40\%), but subsequently reaches pre-aggression levels as the war unfolds. Finally, we can see a clear shift from UA to RU even early on (+129\%). This switch becomes even more striking with the outbreak of the war, as users are actively changing their behaviour such that average user is 250\% more likely to tweet in UA over RU in the span of a single year. \autoref{tab:full_results_lang_model} reports all relevant changes.
%Next, we take a look at behavioural changes using all three language models (see \autoref{fig:smooth_lang}). For RU over EN we see a constant decline of the odds over time to tweet in RU (-33\%), which further speeds up once the aggression starts (-55\%). For UA over EN we see the exact opposite, as over time users are more likely to tweet in UA (+81\% in odds). This change reverses once aggression starts and the war is breaking out (-40\%), but subsequently reaches pre-aggression levels afterwards. Last but not least, we can see a clear shift that users switch from UA to RU even early on (+129\%). This switch becomes even more striking with the outbreak of the war, as users are actively changing their behaviour such that average user is 250\% more likely to post in UA over RU in the span of only 51 weeks. \autoref{tab:results_smooth_lang_model} reports all relevant numbers.
\\

Connecting these language shifts with the results on tweeting activity, we find that the initial decline in EN and RU tweeting activity is not limited to monolingual users. Instead, users are actively shifting towards UA, by reducing their amount of RU and EN tweets (with a stronger shift from RU than EN respectively). Similarly, the temporary increase in EN tweeting behaviour leading up to the war can be linked to both UA and RU users. Finally and most importantly, the decline of RU and the rise of UA tweeting behaviour that manifests with the war is strongly driven by a major language shift (2.5x) from RU to UA.
\\

\begin{comment}
\begin{table}[!htbp] 
\centering
\begin{threeparttable}
\caption{Average Behaviour Change Effect Sizes - Language Models}
\fontsize{9.5}{11.5}\selectfont
\begin{tabular}{c c c c c} 
 \hline
\multirow{2}{*}{Language}  & \multicolumn{4}{c}{Change in Week Interval}   \\ 
\cline{2-5}
&  Start - Aggression  & Aggression  - War   & War - End & Aggression - End   \\
 \hline
UA vs. RU & +130.99\% & +52.08\% & +129.24\% &  +248.63\%\\ 

UA vs. EN & +63.41\% & -33.61\% & +92.663\% &  +27.90\%\\ 

RU vs. EN & -38.89\% & -38.69\% & -20.659\% &  -51.36\%\\ 
\hline
\end{tabular}
\caption*{}
\label{tab:results_smooth_lang_model}
\end{threeparttable}
\end{table}
\end{comment}
%link towards other results
%add other model specifically estimated for non-language switch users
%Overall, we can conclude that there are (minor) sample changes pre-dating any aggression, with more major shifts with the outbreak of the war and the time after. In terms of behaviour, we see mostly steady shifts already before the war, which seem to significantly speed up once aggression starts. Here, we want to specifically emphasize the major behavioural shift from UA to RU (+250\%).\\

\begin{figure}[!htbp] 
     \vspace{-1cm}
     \centering
     \begin{subfigure}[b]{1\textwidth}
         %\hspace*{-1.5cm}
         %\vspace{-1cm}
         \centering
         \includegraphics[scale=0.62]{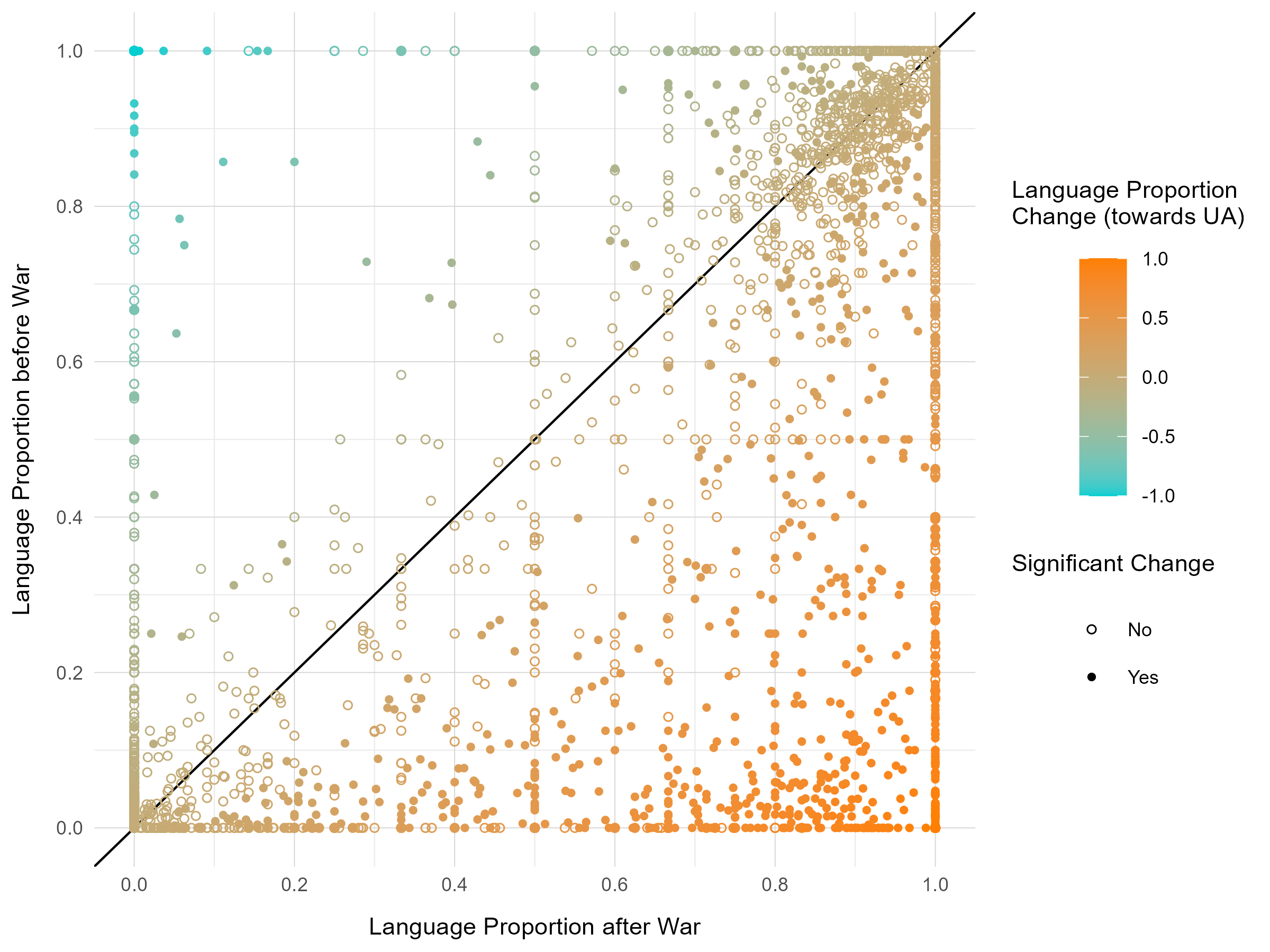}
         \caption{Scatterplot of users' language proportions before and after the outbreak of the war. We are only considering users who tweet in either RU or UA (or both) before and after (n = 3237). The points are colored with respect to each user's shift in language (1 denotes a complete shift to UA, -1 a complete shift to RU, 0 no shift). The straight line through the origin covers all points without a shift. Significant shifts (p<0.05) are denoted through full (non-empty) points. Significance was calculated by individually comparing each user's language proportion through a two-sided z-test before and after war outbreak (24th February 2022). n = 1808 (821 significant) shifts towards Ukrainian, n = 818 (106 significant) shifts towards Russian.}
         \label{fig:scatter_switch}
         \vspace{0.35cm}
     \end{subfigure}
     %\hfill
     \begin{subfigure}[b]{1\textwidth}
      %\hspace*{-0.7cm}
        \centering
         \includegraphics[scale=0.38]{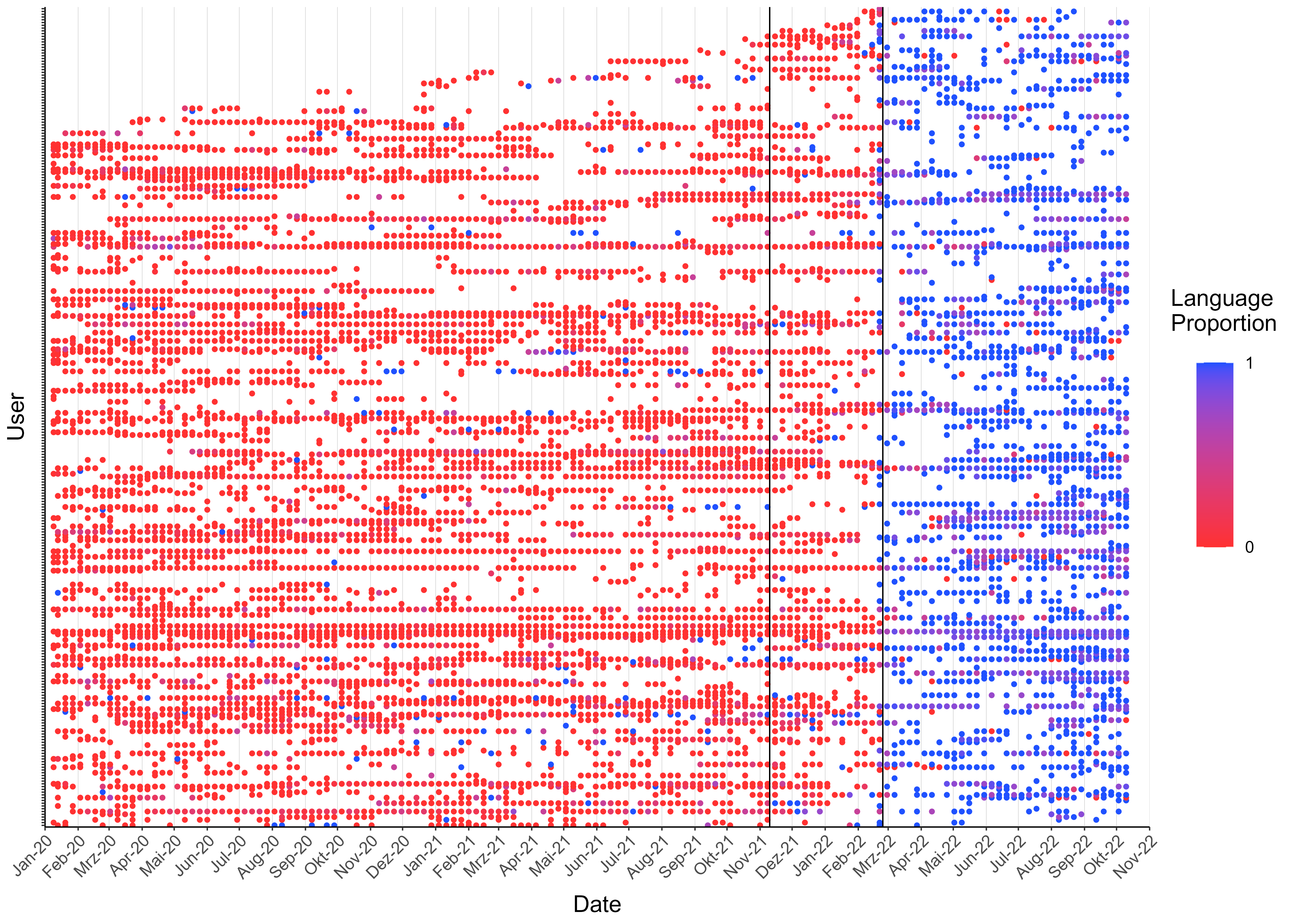}
         \caption{Scatterplot of users' language proportion in each week over time. Each row (on the y-axis) denotes one of the n = 295 users with a statistically significant hard-switch from RU to UA. The points are colored with respect to each user's language proportion in the respective week (145 total weeks). Missing points indicate that a user was not tweeting in the respective week. The first vertical line denotes the mobilization of the Russian troops along the Ukrainian border (11th November 2021). The second line denotes the outbreak of the war (24th February 2022).}
        \label{fig:scatter_switch_time}
     \end{subfigure}
\caption{Language proportion scatterplots of users. The language proportion ranges from $[0,1]$, with 0 being defined as 100\% of a user's tweets being in RU, and 1 as 100\% of tweets in UA. Only RU and UA tweets of each user are considered.}
 \label{fig:scatters}
\end{figure}

We visualize and demonstrate this substantial behavioural language shift from UA to RU in \autoref{fig:scatters}. \autoref{fig:scatter_switch} plots the language proportion of each user (UA to RU; from 0 to 1) that tweet in either language before (y-axis) and after the war (x-axis). Hence, along the straight black line through the origin we have users that do not switch language (top right UA, bottom left RU), users above the line switch to RU, below the line to UA, with users switching completely from one language to the other being located in either the top left (all tweets in UA to all in RU) or bottom right corner. Statistically significant ($p<0.05$) language shifts from before to after war outbreak for each user are marked. From the figure it becomes evident that there are many users that do not switch language (in both UA and RU), as well as many users clearly switching from RU to UA at various levels, whereas there are only very few switching from UA to RU. 
\\

%1756, 672, 487 (72.4\%), 143 (21.3\%)
In this sample of users who tweet in either RU or UA both before and after the outbreak of the war (3237 users), we have 1363 users who predominately tweet in RU (>80\% of tweets) before the war. Of those, 839 (61.6\%) tweet more in UA after the war, with 566 (41.5\%)  reporting a significant behavioural change (p < 0.05). Out of those 850 users, 341 (25\%) even switch to predominately tweeting in UA (>80\% of tweets), i.e. perform a "hard-switch", with 296 (21.7\%) statistically significant hard-switches (p < 0.05). We pick those 296 users and plot their weekly language proportion over time in \autoref{fig:scatter_switch_time}. Red points denote 100\% of the tweets being phrased in RU, blue points denote the same in UA. From the figure, we can clearly see a substantial break and change in behaviour around the time the war breaks out (second black line), as most of the users switch from RU to UA around this mark.
\\

On Ukrainian side, we have 1172 users who predominately tweet in UA (>80\% of tweets) before the war. Of those, 471 (40.2\%) tweet more in RU after the war, with only 83 (7.1\%) reporting a significant behavioural change (p < 0.05). More importantly, we only observe 35 (3\%) hard-switches, out of which 20 (1.7\%) are significant (p < 0.05). Hence, there are only very few UA tweeting users for which we can report a significant switch towards RU after the war.
\\

Finally, we analyze potential differences in those RU users that perform a hard-switch to UA from those that do not. We find that there are significant differences ($p < 0.05$) in the median in various user characteristics between the two groups. Users switching have more followers (+54.5\%), a higher tweet frequency (+47.7\%) as well as a higher like frequency (+48.9\%) and published more Ukraine geo-tagged tweets during the study period (+49.1\%), whereas there are only small non-significant differences in account age (+9.7\%; $p = 0.13$) and followings (+13.8\%; $p = 0.15$). For more information and a full breakdown see section \ref{user_differences}.

\section{Discussion}
%talking to world about cats in Ukraine

We collected geo-tagged tweets from Ukraine and analyzed tweeting activity and language choice before and during the Russian-Ukrainian War from 9th January 2020 to 12th October 2022. Due to the nature of our longitudinal dataset, in which we observe the same set of users across the study period, we were able to disentangle shifts in the user sample, arising from user turnover, from behavioural changes of the actively tweeting users.
\\

%At the beginning of 2020, Russian was the predominant language on Ukrainian Twitter, although survey results from 2017 show that 68.3\% report Ukrainian as their sole native language \citep{kulyk2018shedding}. We interpret this as Twitter either being more popular among the Russian native speaking part of the population, and/or Russian being used by Ukrainian native speakers, as it allows for regional intercountry dialogues. Notably, a large portion of Ukrainian citizens speak at least some Russian at home \citep{kulyk2018shedding}. \\

We find there is a steady long-term shift away from Russian towards Ukrainian already before the war, as the Ukrainian tweet probability rises substantially (vs. Russian; 33\% to 47\%). 
This shift can be largely attributed to behavioural changes. The actively tweeting users reduce their number of Russian tweets in favour of Ukrainian over time. This finding is in line with trends observed over a 20-year period between the 1989 and the last conducted census in 2001 \citep{stebelsky2009ethnic} and more recently across surveys \citep{kulyk2018shedding}, where the share of people reporting Ukrainian as their native language perpetually rose over time. Notably, with the Euromaidan protests and the subsequent Russian military intervention in 2014, this shift seems to have sped up, as citizens ethnonational identification and everyday language use is substantially shifting towards Ukrainian.
\\

The pattern we observe on Ukrainian Twitter is relatively similar. We find a gradual but substantial language shift already pre-war, which drastically accelerates with the start of the Russian aggression in November 2021 and the subsequent outbreak of the war. In the span of a few months, Ukrainian tweet probability rises from 47\% to a remarkable 76\%. While some of this increase can be explained by Russian tweeting users leaving and Ukrainian users joining (+101\% in odds to tweet in Ukrainian), the major factor is a behavioural change (+249\% in odds to tweet in Ukrainian), with a rise in Ukrainian (+56\%) and a decrease in Russian tweeting activity (-20\%). Notably, we show that out of those users predominately tweeting in Russian before the war, roughly half of them tweet more in Ukrainian after. Strikingly, around a quarter of them switch to predominately tweeting in Ukrainian, i.e. performs a hard-switch. It is worth noting, that we do not observe more than a handful of switches in the other direction. This shift from Ukrainian to Russian is in line with recent reports and small-scale surveys outlining the war as the cause for the recent shifts in language use across Ukraine \citep{Guardian.06.03.2023,Multilingual.2022}. Our work confirms these findings on a large-scale on social-media and pinpoints this substantial change exactly to the outbreak of the war.
\\

Russian users that perform a hard-switch to Ukrainian seem to be more active on Twitter and have a larger follower base, despite the overall number of followers being fairly low (median of 119 vs. 77). Nonetheless, we find these differences to be statistically significant. While these would not be deemed as influencer accounts, their behaviour could be attributed to a form of signalling to their user-base of their opposition to the war. %Unfortunately, this is difficult to confirm, as even analyses on the content of the tweets may not be fruitful due to concerns of individual safety stating clear oppositions to the war. 
\\

Furthermore, we find a long-term behavioural shift away from English tweeting activity up until November 2021.
This could be interpreted as a reduction in talking to a broader international audience during that time \citep{smith2015english,christiansen2015rise,moreno2022reexamining}, due to the fact that English is the most widely understood language on the internet by far \citep{statista_2022}. 
However, not surprisingly, with the mobilization of the Russian troops along the Ukrainian border and specifically in the weeks leading up to the war, with a spike during outbreak, we observe a substantial shift towards English, as we hypothesize users wanted to let the world know what was happening and called for aid. While we record a large influx of English speaking users during that time, we can also observe a substantial behavioural shift. Already active users tweet substantially more in English, independent of the language they were normally tweeting in. As the war continues to unfold, this somewhat reverses, with some of the newly joined English users leaving and behaviour reverting, although not to pre-aggression levels. With the world being more aware of the situation, and the international community supporting Ukraine in various ways \citep{european_commission_eu-ukraine_2023,the_white_house_fact_2023}, we hypothesize users have less reasons to continue tweeting in English. Instead, they return back to intra-national discussions and thus their native language(s).
\\

%Our work strictly focused on a large-scale analysis of tweeting activity and language use on Ukrainian Twitter. Hence, 
%Our work shed light on tweeting activity and language use on Ukrainian Twitter before and during the war through a large-scale encompassing analysis. 

We recognize that our study provides a foundation towards a better understanding on how the Ukrainian population reacted to the Russian invasion both on- and offline. Future work could potentially take a closer look on content and sentiment of tweets through multilingual topic modelling and sentiment analyses. This could be augmented through the use of media objects attached to tweets such as images and videos. An investigation of retweet and follower networks could provide additional information on user characteristics as well as interactions in order to find differences between the users that are shifting language compared to those that are not. Naturally, any analysis could be extended to other social media platforms. %Last but not least, offline?
\\

In summary, our work investigated tweeting activity and language choice on Ukrainian Twitter before and during the Russian-Ukrainian War through a large-scale longitudinal study. We observe a substantial shift away from the Russian language to Ukrainian, with more than half of the predominately  Russian-tweeting users shifting towards Ukrainian, and a quarter of them even performing a hard-switch to Ukrainian, as the war broke out. We may interpret this as citizens' increasing opposition to Russia and a return to the country's linguistic roots as well as a push towards a conscious self-definition of being Ukrainian. We deem this a powerful political message to send to a global audience.

\section{Methods}
This study was ethically approved by the ethics commission of the faculty of mathematics, computer science and statistics at Ludwig-Maximilians-Universität (LMU) München, Germany. The reference identifier is EK-MIS-2022-127.

\subsection{Data Collection} \label{data_collection}
The original Twitter dataset obtained from the 1\% stream consisted of 4,102,982 tweets (see section \ref{results_data} for details). As we began cleaning, we noticed gaps with missing tweets, most likely due to server and internet outages during the real-time data collection process. Hence, we retrospectively identified and filled all gaps. To do this, we first identified all time windows $>10$ min without any tweet and added them to our download queue. Days with more than two of such time windows were added to the queue as a whole. We then queried the Twitter Research API 2.0 using the \emph{tweets/search/all} endpoint to obtain tweets with Ukrainian geoinformation for all time windows in this queue and added the newly obtained tweets to our original dataset. Finally, we repeated this process for the 15 days with the least amount of tweets in our dataset. After removing all duplicates, this meant we added a total of 350,359 additional tweets to our dataset this way. We perform our sensitivity analysis (see section \ref{sensitivity_analysis}) after this step. We clean this dataset by removing spam as well as potential spam bots and accounts, as described in section \ref{cleaning_and_preprocessing}.

\subsection{Sensitivity Analysis} \label{sensitivity_analysis}
After the collection of tweets as described in section \ref{data_collection}, we evaluate the completeness of the dataset, i.e. if we were able to recover most of the tweets published in Ukraine during that time, using the following strategy. We draw a random subset of 29 days from our analysis period and draw tweets from the Twitter Research API 2.0 using the \emph{tweets/search/all} endpoint, which returns all historic tweets that have not been deleted since. We report a coverage of 98.24\% (SD: 3.09\%). More importantly, in the opposite direction we are only able to report a coverage of 77.67\% (SD: 9.55\%). Hence, employing our strategy using the real-time stream offers substantially more tweets, which have been deleted since (for more information on tweet deletion and its effects see \cite{pfeffer2022sample}). Moreover, this suggests we were able to recover most of the geo-tagged tweets from Ukraine using our strategy.

\subsection{Data Cleaning \& Pre-processing} \label{cleaning_and_preprocessing}
For cleaning our dataset, we first train a Twitter bot detection model using a random forest (RF), as described in \cite{yang2020scalable}. We use the exact same model as described in the authors' work (except for removing the attribute \emph{profile\_use\_background\_image}, which is no longer available from the Twitter API), using the training datasets \emph{botometer-feedback}, \emph{celebrity}, \emph{political-bots}, as well as 100 manually labelled Twitter accounts from our dataset. To evaluate performance, we first set up a nested cross validation (CV) routine, with both a 5-fold CV in the inner and outer loop. The inner CV is used for hyperparameter tuning, tuning both the number of trees as well as the minimum node size of the RF, whereas the outer loop is used for evaluating model performance. This results in an average area under the receicer operator characteristics curve (AUROC) of 0.9837 and an average area under the precision-recall curve (AUPRC) of 0.7707. For our final model, we replicate this procedure, by setting up a 5-fold CV on the entire dataset to find the best performing hyperparameters. We then train our RF on the entire dataset and use this model to identify bots and spam accounts in our dataset. 
\\

As we are only interested in removing the most prevalent spam, %and at the same time there is a large influx of new Twitter users after the outbreak of the war, 
we opt for a conservative removal strategy to not falsely remove too many real and non-spam users. Hence, we only remove users with a predicted bot probability $>50\%$ and more than 10 tweets since account creation as well as users with a predicted bot probability $>30\%$ and more than 10,000 tweets. While thresholds of 50\% and 30\% respectively might not seem conservative, in the given setting, in which the bot class is heavily underrepresented (3.7\% of observations in training dataset), an F1-optimizing threshold on the training dataset would lie far below that. We are somewhat less conservative with users that published over 10000 tweets, as in most cases they are spam accounts (e.g. related to bitcoins or NFTs). We do to not remove users with less than 11 tweets, as even for a human it becomes incredibly difficult to determine if a user is a bot with such limited amount of information to draw from. At the same time, we noticed a large influx of new users after the outbreak of the war who exclusively called for help in a short span of time, a behaviour which can easily be mistaken for a bot.
Notably, we do not tune the optimal classification threshold, as the outbreak of the war in Ukraine represents an unprecedented event, with an unusual amount of new users joining (see section \ref{user_activity}). Hence, we expect the distribution between the target label (bot or human) and our features to be different between the bot training dataset and our Ukrainian dataset. Unfortunately, an extensive manual labelling strategy and more elaborate bot detection is beyond the scope of this work and would warrant its own paper. In summary, with this strategy we remove a total of 2021 users and their tweets from our dataset. 
\\

To further identify and remove potential spam accounts, we identify all accounts with more than 100 tweets on a single day (the mean is $\sim 4.4$ and the median = $2$), and remove those 257 users from the dataset. We also noticed an unusual amount of Tweets containing the word "BTS" (45,579; referring to the Korean K-Pop band, see \cite{lee2020music} for more information) with spikes on specific days, which we subsequently filter out. Next, we identify and remove any tweets published by the same user that contain the exact same text as their previous tweet if both tweets were published within a one minute window. Fifth and finally, we filter out any tweets with the \emph{source} attribute not being equal to Instagram or Twitter. That way, we discard any tweets automatically published by social media schedulers such as \emph{dlvr}, which are often used by news agencies or other companies.

\subsection{Tweet Modelling} \label{tweet_modelling}
We define the number of tweets $Y_{t,u,l}$ made in week $t$ by user $u$ in language $l$. As tweets are count data, we model the $Y_{t,u,l}$ to follow a Poisson distribution with intensity $\lambda_{t,u,l}$, where
\begin{equation*}
\lambda_{t,u,l} = exp(\mu + s_l(t) + W_{u,l}).
\end{equation*}
Here, $\mu$ is a general time-constant intercept, which captures the average tweet intensity over all users, languages and weeks. The $W_{u,l}$ are language-specific time-constant random intercepts for each user $u$, assumed to be normally distributed. They capture by how much the average tweeting behaviour (more or less tweets) of each user in each language differs from the general mean $\mu$. Finally, $s_l(t)$ denotes a smooth global time trend for each language $l$ (Ukrainian, Russian, English) and captures changes in the tweeting behaviour over all users over time. Hence, with the latter, we can measure behavioural changes of the users over time (e.g. are users tweeting more with the outbreak of the war?), whereas the random intercepts measure changes in the user sample over time (e.g. are users that enter the platform after the war tweeting more on average?). We fit the model with the R package \emph{mgcv} v1.8.41 \citep{wood2017generalized} using the GAM implementation for very large datasets \emph{bam}. To speed up the estimation, we use the discrete option, which discretizes covariates to ease storage and increase efficiency. For fitting $s_l(t)$, we employ thin plate regression splines. Our estimation sample consists of $y$ = 1,045,245 observations, with $t$ = 143 weeks, $l$ = 3 languages and $u$ = 13,643 users. For our fitted model, we report an explained deviance of 71.3\%.
\\

The effect sizes in the main text are calculated as follows. For the behavioural effects we derive the change in $s_l(t)$ between two respective dates $t_1$ and $t_2$ and take the $exp(.)$, i.e. $exp(s_l(t_2)-s_l(t_1))$ for each language $l$. The result is the change in expected tweeting activity due to behavioural changes, when controlling for the in- and outflux of users. The sample effects are derived by averaging the random effects of the active users at the two respective dates and taking the $exp(.)$, i.e. $exp(\overline{W}_{t_2,l} - \overline{W}_{t_1,l})$. We define $\overline{W}_{t,l}$ as the average random effect in language $l$ over all users $u$ active at time point $t$. This captures the averaged change in expected tweeting activity due to a change in average tweeting intensity of the active users, when controlling for behavioural changes.
\begin{comment}
 \begin{figure}
     %\hspace*{-1.5cm}
        \centering
         \includegraphics[scale=0.36]{Model_Figures/violin_res_tweet_model.png}
         \caption{Violin plots at the 6-weekly aggregate level across languages. The first row denotes Ukrainian, the middle Russian, and bottom, English.}
        \label{fig:violin_res_tweet_model}
     \end{figure}
\end{comment}

\subsection{Language Modelling} \label{language_modelling}
To model users' pairwise language probability, we refrain from a multinomial modelling strategy, as even with a weekly setup our dataset is particularly large. (To the best of our knowledge, a package with a parallel estimation routine for large datasets that can fit a GAMM for a multinomial distribution does not exist.) Instead, we model each pairwise probability separately through a binomial distribution. Our pairwise evaluation gives us a total of three different language pairs (UA over RU, UA over EN, RU over EN), for which we model the probability $\pi$ to tweet in language one (subsequently $l_1$) over language two (subsequently $l_2$). The order in which we specify these pairs is irrelevant, as the probability to tweet in $l_2$ over $l_1$ is simply $1 - \pi$. More specifically, we define $X_{t,u}$ as the number of tweets made in week $t$ by user $u$ in $l_1$. We assume $X_{t,u} \sim Binomial(n_{t,u},\pi_{t,u})$, where $n_{t,u}$ denotes the total number of tweets made by user $u$ in week $t$ (sum of tweets in $l_1$ and $l_2$) and $\pi_{t,u}$ corresponds to the probability to tweet in $l_1$ over $l_2$. We assume that $n_{t,u}$ is known and instead model $\pi_{t,u}$ by setting
\begin{equation*}
\pi_{t,u} = f(\mu + s(t) + W_{u}),
\end{equation*}
where $f(.)$ is defined as the logistic function. Similarly to before, $\mu$ is a general time-constant intercept, which captures the average mean probability over all users and weeks to tweet in $l_1$ over $l_2$. Again, the $W_{u}$ are time-constant random intercepts for each user $u$ that capture by how much the average probability differs from the general mean $\mu$, and are assumed to be normally distributed. The smooth global time trend $s(t)$ captures changes in the probability over all users over time. Hence, as before, we can measure behavioural changes of the users over time with the latter (are users actively changing the language they are tweeting in?), whereas the random intercepts measure changes in the sample over time (how does the language probability of users entering/leaving the platform evolve?). We estimate this model specification for all three aforementioned language-pairs with the R package \emph{mgcv} v1.8.41 \citep{wood2017generalized} using the GAM implementation for very large datasets \emph{bam}. To speed up the estimation, we use the discrete option, which discretizes covariates to ease storage and increase efficiency. For fitting $s(t)$, we employ thin plate regression splines. Users not tweeting in either of the two languages of the respective language pair, need to be discarded by definition. Hence, for UA over RU our estimation sample consists of of $x$ = 194,178 observations, with $t$ = 143 weeks and $u$ = 10,531 users. For UA over EN: $x$ = 146,984, $t$ = 143, $u$ = 9,133. For RU over EN: $x$ = 170,853, $t$ = 143, $u$ = 10777. For our fitted models, we report explained deviances of: 85.8\% (UA over RU), 90.5\% (UA over EN) and 90\% (RU over EN).
\\

The coefficients of a logistic regression, as employed here, must be interpreted with respect to changes in the odds (also known as odds ratio). The odds ratio is defined as $odds = p/(1-p)$. Hence, it describes how likely an event is going to happen compared to not happen. In this setting, it describes how likely it is to tweet in language 1 over language 2.
\\

The effect sizes in the main text are calculated as follows. For the behavioural effects we derive the change in $s(t)$ between two respective dates $t_1$ and $t_2$ and take the $exp(.)$, i.e. $exp(s(t_2)-s(t_1))$ for each of the three models. The result is the change in odds to tweet in $l_1$ over $l_2$ due to behavioural changes, when controlling for the in- and outflux of users. The sample effects are derived by averaging the random effects of the active users at the two respective dates and taking the $exp(.)$, i.e. $exp(\overline{W}_{t_2} - \overline{W}_{t_1})$ for each of the three models. We define $\overline{W}_{t}$ as the average random effect over all users $u$ active at time point $t$. This captures the averaged change in odds due to a change in average tweeting probability of the active users, when controlling for behavioural changes.

\clearpage
\section{Extended Data}

\subsection{Language Distribution} \label{language_distribution}

\begin{figure}[!htbp] 
    %\vspace{-2.5cm}
    \centering
    \includegraphics[scale=0.6]{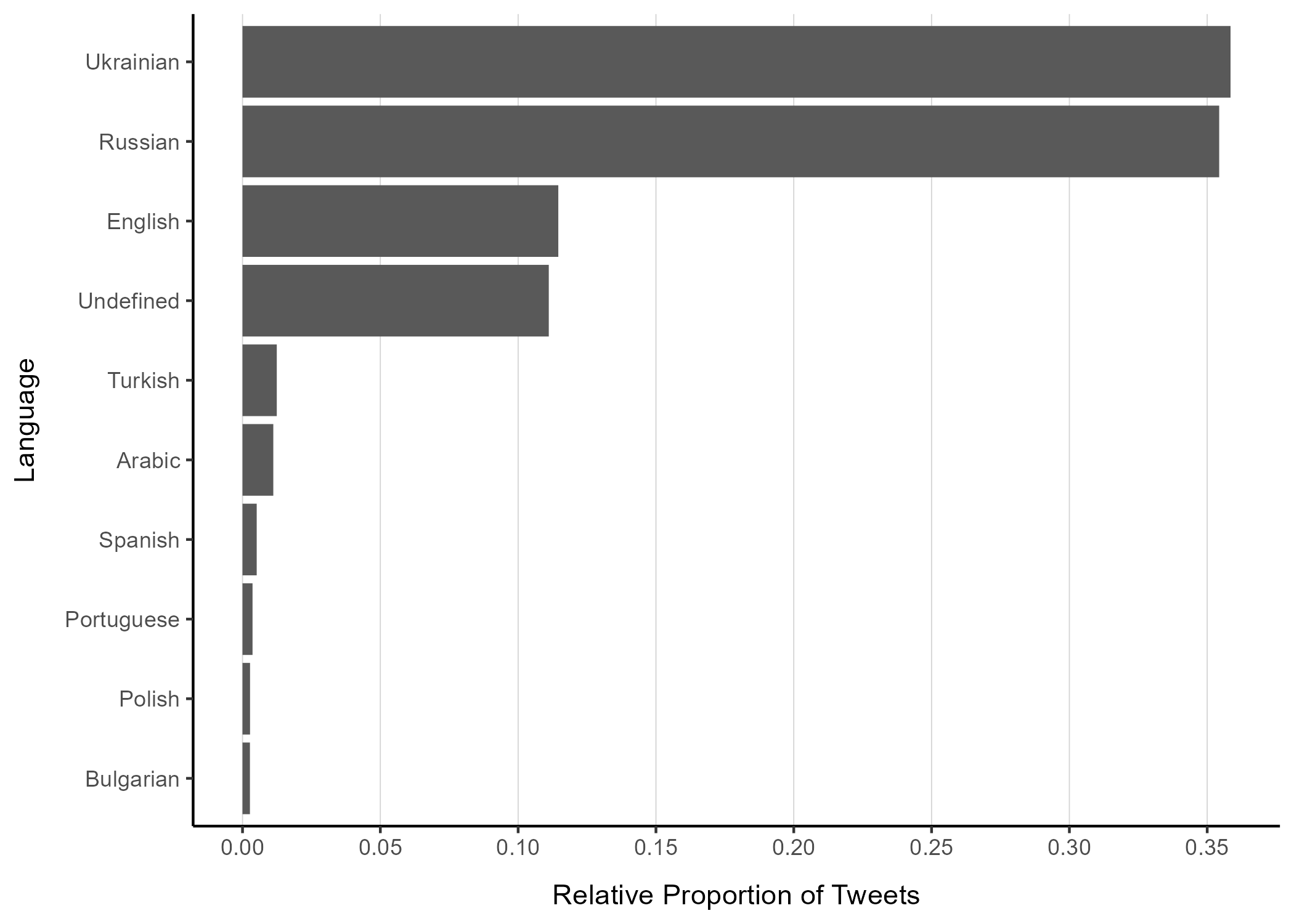}
    \caption{Relative distribution of the top 10 languages across the entire sample after preprocessing and cleaning (n = 2,845,670 tweets). "Undefined" consists of tweets that are too short, contain only hashtags, contain only mentions or only have media (links), for all of which a language is not available.}
    \label{fig:lang_distr}
\end{figure}

\subsection{Differences in User Characteristics for Russian Users} \label{user_differences}
We evaluate differences in user characteristics between the 1,363 user who predominately tweet in Russian (>80\% of tweets) with respect to their language shift with the outbreak of the war in \autoref{tab:user_differences}. Column 2 reports the median of the respective user characteristic for those 1067 Russian users that do not perform a statistically significant ($p<0.05$) hard-switch to Ukrainian (>80\% of tweets) with the outbreak of the war, column 3 for the 296 users that do. To determine significance, we employ a two-sided z-test on each user's language proportion (\% tweets in UA) before and after the outbreak of the war. Column 4 reports the relative difference from the switch group to the no switch group, with bold values indicating significant differences between the two groups ($p<0.05$). Column 5 reports the p-value of the two-sided statistical significance test on the difference in median between the two groups using a chi-squared test. Column 6 the chi-squared statistic.
\begin{table}[H]
\centering
\begin{threeparttable}
\caption{Median \% Differences in User Characteristics}
\fontsize{10}{12}\selectfont
\begin{tabular}{l l l | l l l} 
 \hline
 User Characteristic & No Switch & Switch & Difference & P-Value & $\rchi^2$\\
 \hline
 Followers & 77 &  119 & \textbf{+123.61\%}  & 0.004 & 8.223\\ 

 Tweet Frequency & 0.79 &  1.16 & \textbf{+47.73\%} & 0.021 & 5.352\\

 Like Frequency & 0.84 & 1.25 & \textbf{+48.93\%}  & 0.021 & 5.352\\

 \# of Tweets in Ukraine & 57 & 85 & \textbf{+49.12\%}  & 0.001 & 10.639\\

 Account Age (Month) & 98.28 & 107.84 & +9.73\%  & 0.127 & 2.326\\

 Followings & 116 & 132 & +13.9\% & 0.155 & 2.023\\

 \hline
\end{tabular}
\caption*{Notes: n = 1,067 users in the no switch group, n = 296 users in the switch group. Followers are the number of accounts that follow a user. The tweet frequency reports the number of tweets per day. The like frequency the number of liked tweets (by the user) per day. "\# of tweets in Ukraine" reports the number of tweets in our dataset. The account age reports the number of months a user account has existed from account creation to their latest tweet in our dataset. Followings report the number of accounts a user is following. All user characteristics (except \# tweets in Ukraine) are derived from the Twitter API, using the provided fields accompanying the user's latest tweets.}
\label{tab:user_differences}
\end{threeparttable}
\end{table}

\section*{Funding Statement}
\noindent This work is supported by the Helmholtz Association under the joint research school ‘‘Munich School for Data Science - MUDS’’. This work is also supported by the European Research Council (ERC) under the European Union's Horizon 2020 research and innovation programme (grant agreement No. [ERC-2016-StG-714087], Acronym: \textit{So2Sat})

\bibliographystyle{elsarticle-harv} 
\bibliography{bib}

\end{document}